\documentclass[10pt,letterpaper]{article}
\usepackage[top=0.85in,left=2.75in,footskip=0.75in]{geometry}

\usepackage{amsmath,amssymb}

\usepackage{changepage}

\usepackage{textcomp,marvosym}

\usepackage{cite}

\usepackage{nameref,hyperref}

\usepackage[right]{lineno}

\usepackage[nopatch=eqnum]{microtype}
\DisableLigatures[f]{encoding = *, family = * }

\usepackage[table]{xcolor}

\usepackage{array}

\newcolumntype{+}{!{\vrule width 2pt}}

\newlength\savedwidth

\raggedright
\usepackage[aboveskip=1pt,labelfont=bf,labelsep=period,justification=raggedright,singlelinecheck=off]{caption}

\usepackage{float}
\usepackage{wrapfig}

\newcommand{\approptoinn}[2]{\mathrel{\vcenter{
  \offinterlineskip\halign{\hfil$##$\cr
    #1\propto\cr\noalign{\kern2pt}#1\sim\cr\noalign{\kern-2pt}}}}}

\usepackage{MnSymbol}

\usepackage{amsthm}
\usepackage{amsmath}
\DeclareMathOperator*{\argtopk}{arg\,top\,k}
\DeclareMathOperator*{\argmax}{arg\,max}

\usepackage{multirow}

\usepackage[noend,linesnumbered,ruled]{algorithm2e}

\theoremstyle{definition}
\newtheorem{definition}{Definition}[section]

\usepackage[most]{tcolorbox}
\newtcolorbox{highlighted}{colback=yellow,breakable}

\usepackage{lineno}

\makeatletter
\renewcommand{\@biblabel}[1]{\quad#1.}
\makeatother

\usepackage{lastpage,fancyhdr,graphicx}
\usepackage{epstopdf}
\fancyheadoffset[L]{2.25in}
\fancyfootoffset[L]{2.25in}
\begin{document}
\vspace*{0.2in}

\begin{flushleft}
{\Large
\textbf\newline{Orchard: building large cancer phylogenies using stochastic combinatorial search} 
}
\newline
\\
Ethan Kulman\textsuperscript{1,2},
Rui Kuang\textsuperscript{2},
Quaid Morris*\textsuperscript{1},
\\
\bigskip
\textbf{1} Computational and Systems Biology Program, Sloan Kettering Institute, New York, NY, USA
\\
\textbf{2} Department of Computer Science and Engineering, University of Minnesota, Minneapolis, MN, USA
\\
\bigskip

%
%





* MorrisQ@mskcc.org

\end{flushleft}
\section*{Abstract}
Phylogenies depicting the evolutionary history of genetically heterogeneous subpopulations of cells from the same cancer, i.e., \textit{cancer phylogenies}, offer valuable insights about cancer development and guide treatment strategies. Many methods exist that reconstruct cancer phylogenies using point mutations detected with bulk DNA sequencing. However, these methods become inaccurate when reconstructing phylogenies with more than 30 mutations, or, in some cases, fail to recover a phylogeny altogether.

 Here, we introduce \textit{Orchard}, a cancer phylogeny reconstruction algorithm that is fast and accurate using up to 1000 mutations. Orchard samples without replacement from a factorized approximation of the posterior distribution over phylogenies, a novel result derived in this paper. Each factor in this approximate posterior corresponds to a conditional distribution for adding a new mutation to a partially built phylogeny. Orchard optimizes each factor sequentially, generating a sequence of incrementally larger phylogenies that ultimately culminate in a complete tree containing all mutations.

Our evaluations demonstrate that Orchard outperforms state-of-the-art cancer phylogeny reconstruction methods in reconstructing more plausible phylogenies across 90 simulated cancers and 14 B-progenitor acute lymphoblastic leukemias (B-ALLs). Remarkably, Orchard accurately reconstructs cancer phylogenies using up to 1,000 mutations. Additionally, we demonstrate that the large and accurate phylogenies reconstructed by Orchard are useful for identifying patterns of somatic mutations and genetic variations among distinct cancer cell subpopulations.

\section*{Author summary}

Somatic mutations accumulate during cancer progression, leading to genetically distinct subpopulations of cancer cells. Understanding the evolutionary relationships among these subpopulations reveals critical events and offers clinical insights for treatment. We developed a novel machine learning algorithm, Orchard, which uses point mutations detected via bulk DNA sequencing to reconstruct phylogenies depicting the genetic evolution of an individual cancer. We evaluated Orchard on 90 simulated cancers and 14 B-progenitor acute lymphoblastic leukemias. Our results demonstrate that Orchard is fast and accurate at reconstructing phylogenies with up to 1,000 mutations detected in 100 cancer samples. Orchard's ability to reconstruct large phylogenies motivated a novel approach: applying agglomerative clustering to these trees to identify clusters of mutations that co-occur in cells. Our results show that this new clustering approach accurately identifies mutation clusters in real data, often outperforming state-of-the-art methods.

\section*{Introduction}
Cancerous cells contain somatic mutations that override normal cellular controls \cite{hanahan_hallmarks_2000, nowell_clonal_1976}. Evolution progresses as cancer cells acquire additional mutations that provide selection advantages \cite{greaves_clonal_2012}, leading to genetically distinct cancerous cell subpopulations (i.e., \textit{subclones}) characterized by sets of shared mutations \cite{gerstung_evolutionary_2020}. Phylogenies describing the ancestral relationships of subclones in an individual cancer can reveal important evolutionary events, help to characterize the cancer's heterogeneity \cite{dentro_characterizing_2021} and progression \cite{hirsch_genetic_2016, kent_order_2017}, support the discovery of new driver mutations \cite{koch_using_2022}, and provide clinical insight for treatment regimens \cite{pogrebniak_harnessing_2018}. Many cancer evolution studies use bulk DNA from multiple cancer samples to reconstruct phylogenies depicting the evolution of subclones in an individual cancer \cite{adams_consensus_1972,nik-zainal_life_2012,jamal-hanjani_tracking_2017,light_germline_2023,sakamoto_evolutionary_2020,schuh_monitoring_2012,villani_clinical_2022}. These phylogenies are also prerequisites for inferring migration histories of metastases \cite{el-kebir_inferring_2018, turajlic_deterministic_2018}.

A \textit{mutation tree} is an arborescence that can represent portions of a cancer phylogeny that satisfy the infinite sites assumption (ISA). The ISA posits that each mutation is acquired only once and is never reverted. A phylogeny that adheres to the ISA is a \textit{perfect phylogeny}; cancer phylogenies are often perfect because of low mutation rates ($\sim$1-10 mutations / megabase) \cite{chalmers_analysis_2017}. By definition, a mutation tree is a perfect phylogeny, and each non-root node in a mutation tree represents a somatic mutation present in one or more cancerous cell subpopulations. Bulk DNA data obtained by sequencing multiple tissue samples from the same cancer can be used to reconstruct a mutation tree by solving the \textit{mixed sample perfect phylogeny} (MSPP) problem.

In the MSPP problem, each cancer sample is an unknown mixture of clonal genotypes, i.e, \textit{clones}, that are related to one another via a perfect phylogeny. The input to the MSPP problem is a matrix of the frequencies of each mutation in each sample; these frequencies can be estimated from the variant allele frequencies (VAFs) measured by short read sequencing of the bulk samples. Although perfect phylogeny reconstruction has a linear time algorithm \cite{gusfield_efficient_1991}, MSPP reconstruction is NP-complete even if the mutation frequency estimates are noise-free \cite{el-kebir_reconstruction_2015}. There are some special cases where the noise-free MSPP problem can be solved in quadratic time \cite{sundermann_reconstructing_2021, eskin_efficient_2003, gusfield_haplotyping_2002}.

Until recently, no MSPP reconstruction algorithm could reliably build mutation trees with more than 10 mutations \cite{wintersinger_reconstructing_2022}.
State-of-the-art MSPP reconstruction algorithms such as Pairtree \cite{wintersinger_reconstructing_2022} and CALDER \cite{myers_calder_2019} are capable of accurately reconstructing 30-node mutation trees, and in some rare cases, 100-node mutation trees.
However, most MSPP problems have more than 30 mutations, and this necessitates that mutations are pre-clustered based on their VAFs prior to MSPP reconstruction \cite{caravagna_subclonal_2020,gillis_pyclone-vi_2020,miller_sciclone_2014}. When mutations are clustered together, the reconstructed tree is known as a \textit{clone tree}, and many MSPP algorithms treat each pre-defined cluster of mutations as a single mutation during reconstruction. Clustering mutations is often referred to as subclonal reconstruction, because these mutation sets uniquely define \textit{subclones} which are subpopulations of cancer cells that correspond to clades in the phylogeny. Unfortunately, using mutation VAFs to define subclones is an error-prone process that can lead to incorrectly estimating the number of subclones \cite{salcedo_community_2020, sun_between-region_2017}, and incorrectly grouping mutations together that are in unrelated clones \cite{williams_identification_2016}. Reconstructing a clone tree using poorly defined subclones can yield inaccurate trees and lead to incorrect conclusions about the evolution of a cancer \cite{sun_between-region_2017, williams_identification_2016, satas_tumor_2017}. Pre-defining subclones before tree reconstruction can also result in the loss of critical information such as the partial order of mutation acquisition, which is crucial for analyzing somatic mutation patterns and determining prognosis and treatment strategies for certain hematological cancers \cite{hirsch_genetic_2016, kent_order_2017, benard_clonal_2021, ortmann_effect_2015, schwede_mutation_2024}.


Here, we introduce a new MSPP reconstruction algorithm, Orchard, which reliably reconstructs extremely large mutation trees, alleviating the need to pre-cluster mutations. Orchard accurately solves MSPP problems with as many as 1000 mutations and 100 samples. We benchmark Orchard's reconstructions against Pairtree and CALDER on 14 B-progenitor acute lymphoblastic leukemias (B-ALLs) and 90 simulated cancers. These results demonstrate that: $(1)$ Orchard matches or outperforms Pairtree and CALDER on reconstruction problems with 30 or fewer mutations while being 5-10x faster, and $(2)$ Orchard greatly outperforms both competing methods on reconstruction problems with more than 30 mutations. Orchard achieves greater accuracy and scalability for large reconstruction problems by sampling from a factorized approximation of the mutation tree posterior, a novel result derived in this paper. Each factor in this approximate posterior represents a conditional distribution for adding a new mutation to a partially built tree. Consequently, Orchard samples from this approximate posterior by generating sequences of progressively larger trees, culminating in a complete tree containing all mutations. Orchard uses Gumbel Soft-Max Tricks \cite{kool_stochastic_2019, kool_ancestral_2020, paulus_gradient_2021} to sample trees for each factor, enabling it to sample complete mutation trees without replacement. This approach differs from existing methods, such as Pairtree \cite{wintersinger_reconstructing_2022}, which uses Markov Chain Monte Carlo (MCMC) to explore the space of complete mutation trees, or CALDER \cite{myers_calder_2019}, which formulates the MSPP problem as a mixed integer linear program (MILP). These existing algorithmic approaches have critical pitfalls: MCMC-based methods often converge slowly and are prone to becoming stuck in local maxima because samples are highly auto-correlated, while MILP-based methods can be extremely slow, may fail to recover a solution, or may only be able to produce a single solution in a reasonable amount of time. Orchard avoids these pitfalls because of two crucial properties of the algorithm: it samples without replacement, which mitigates the risk of becoming stuck in local maxima, and its search strategy retains and grows partial tree structures that adhere closely to the ISA, effectively pruning large portions of the search space. 

Orchard's ability to accurately reconstruct large mutation trees enables a novel strategy to flexibly define subclones.
To illustrate this new strategy, we introduce a simple \textit{phylogeny-aware} clustering algorithm that infers subclones from a reconstructed mutation tree. On real data, this simple algorithm recovers expert-defined subclones as well as, and often better, than state-of-the-art VAF based clustering algorithms \cite{caravagna_subclonal_2020,gillis_pyclone-vi_2020,miller_sciclone_2014}. Orchard, along with the phylogeny-aware clustering algorithm, is accessible for free under the MIT License at \url{http://www.github.com/morrislab/orchard}.

\section*{Materials and methods}
\newcommand{\va}{{\mathbf{a}}}
\newcommand{\vd}{{\mathbf{d}}}

Orchard reconstructs cancer phylogenies using point mutations detected via bulk DNA sequencing. Below, we first describe the mixed sample perfect phylogeny (MSPP) problem and its definition using variant allele frequency data. Next, we derive a novel factorized approximation to the mutation tree posterior and demonstrate how to use Gumbel Soft-Max Tricks \cite{kool_stochastic_2019, kool_ancestral_2020} to sample without replacement from it. We then present the Orchard algorithm's pseudo-code and discuss implementation-specific details. Lastly, we describe our phylogeny-aware clustering algorithm that infers clones and clone trees using a mutation tree.
\subsection{The cancer-specific mixed sample perfect phylogeny problem}\label{sec:methods-overview}

The input to a mixed sample perfect phylogeny problem is a matrix of \textit{mutation frequencies}, $F \in \mathbb{R}^{n \times m}$, and the goal is to infer a perfect-phylogeny-compatible binary genotype matrix $B \in \{0,1\}_{n \times n}$. Each column $B_{:v}$ in $B$ represents the binary genotype for clone $v$. The rows of $B$ represent mutations, where if $B_{jv} = 1$, then clone $v$'s genome contains mutation $j$. The rows of $F$ represent mutations and the columns represent samples. Each entry $F_{js}$ is the frequency of mutation $j$ in sample $s$, where $F_{js} \in [0,1]$. Since multiple clones may harbor mutation $j$, $F_{js}$ equals the sum of the frequency of those clones in sample $s$, i.e., $F_{js} = \sum_v B_{jv}U_{vs}$, where  $U_{vs} \geq 0$ and $\sum_v U_{vs} \leq 1$. The mixing coefficients, $U_{vs}$, represent the proportion of cells in samples $s$ with the clonal genotype $v$, i.e., the \textit{clonal proportions}. The general relationship between $F$, $B$, and $U$, is: 

\begin{equation}\label{eq:F-equals-BU}
    F=BU.
\end{equation}
The basic MSPP problem assumes that $F$ is noise free, and solving it requires finding a perfect-phylogeny-compatible binary genotype matrix $B$ and a \textit{clonal proportion matrix} $U \in \mathbb{R}^{n \times m}$ that satisfy Eq \ref{eq:F-equals-BU}.

In the cancer-specific, noisy MSPP problem, we assume that $F$ cannot be observed directly, and instead, we are provided variant allele frequency data $D$, from which a noisy estimate $\widehat{F}$ of $F$ can be derived. The goal is to find $B$ and $U$ that admit an $F$ with the highest possible data fit to $D$. The next section describes how we define a noisy MSPP problem, one step of which is estimating $\widehat{F}$ using the observed bulk DNA data. After that, we formally define a mutation tree.

\subsection{The input to Orchard}\label{sec:orchard-inputs}

The inputs to Orchard include VAFs for $n$ point mutations in $m$ samples from the same cancer. For each point mutation $j$, Orchard requires a set of tuples $X_j = \big\{\left(a_{js}, b_{js}, \omega_{js}\right)\big\}_{s=1}^m$, where $s$ indexes the bulk cancer sample. The entire data set for all $n$ mutations is represented by a set $D = \left\{X_j\right\}_{j=1}^n$. 

Each tuple $\left(a_{js}, b_{js}, \omega_{js}\right) \in X_j$ represents the bulk DNA data from sample $s$ for the genomic locus containing mutation $j$: $b_{js}$ is the count of reads containing the variant allele $j$; $a_{js}$ is the count of reads containing the reference allele; and $\omega_{js}$ is the \textit{variant read probability}, which is the estimated proportion of alleles per cancer cell that contain the variant allele $j$. The latter value, $0 \leq \omega_{js} \leq 1$, is required to convert from VAFs to mutation frequencies; and the dependence on $s$ permits modelling of sample-specific copy number aberrations (CNAs). Under the ISA, a mutation $j$ initially only affects one allele, so in diploid and haploid regions free of CNAs, $\omega_{js} = \frac{1}{2}$ and $\omega_{js} = 1$, respectively. In some cases, $\omega_{js}$ can be estimated by performing an allele-specific copy number analysis \cite{tarabichi_practical_2021}. If mutations are heavily impacted by CNAs, Orchard requires reliable sample specific copy number estimates to produce accurate reconstructions; if these are unavailable, mutations affected by CNAs should be omitted from the inputs provided to Orchard. 

The observed VAF for mutation $j$ in sample $s$, denoted by $\widehat{\lambda}_{js}$, is given by $\widehat{\lambda}_{js} = \frac{b_{js}}{b_{js} + a_{js}}$. This VAF can be converted to an observed mutation frequency,  $\widehat{F}_{js}$, using the following formula: 

\begin{equation}\label{eq:vaf-translation}
    \widehat{F}_{js} = \widehat{\lambda}_{js} / \omega_{js}.
\end{equation}
 The value of $\widehat{F}_{js}$ is an estimate of the percentage of cells in sample $s$ that have mutation $j$.


\subsection{Mutation tree representations of perfect phylogenies}

Orchard represents perfect phylogenies of cancer clones using a mutation tree.
A mutation tree is a perfect phylogeny represented as a rooted, directed tree, i.e., an arborescence. We denote it with $t = \left\{V,E,M\right\}$, where $V$ is a set of nodes, indexed by $v \in \{r\} \cup \{1, 2, \dots, n\}$, with $r$ indicating the root node representing the germline; $E$ is a set of directed edges defining the child/parent relationships among the nodes; and $M$ is a vector of mutations, one for each non-root node. Each node $v \in V \setminus \{r\}$ defines a genetically distinct cell population or \textit{clone} whose clone-defining mutation is $M_v$. Each directed edge $(v,u) \in E$ indicates that $v$ is a parent of $u$, and that $u$ inherits the mutation $M_v$, along with all mutations that $v$ inherited from its ancestors. One implication of this inheritance is that the set of clones in the subtree rooted at $v$ are uniquely defined by having the mutation $M_v$, thus $M_v$ also defines a subclone rooted at $v$. For convenience, we will assign the same index to a clone, its clone-defining mutation, and the node associated with the clone, so clone $v$ is represented by mutation $v$ and $M_v = v$.

Orchard also considers \textit{partial mutation trees}, denoted as $t^{(\ell)} = \left\{V^{(\ell)}, E^{(\ell)}, M^{(\ell)}\right\}$, where the superscript $\ell$ denotes the number of mutations the tree contains. A tree containing only the root node and a single mutation is denoted as $t^{(1)}$, and a \textit{complete mutation tree} containing all $n$ mutations and the root node is denoted as $t^{(n)}$. If $0 < \ell < n$, then $V^{(\ell)} \subset V$ and only the entries $M_v$ for $v \in V^{(\ell)} \setminus \left\{r\right\}$ are defined. To complete a partial tree $t^{(\ell)}$, the remaining mutations $V \setminus V^{(\ell)}$ need to be added to $t^{(\ell)}$. Adding $v \in V \setminus V^{(\ell)}$ to $t^{(\ell)}$ results in a new tree $t^{(\ell+1)} = \left\{V^{(\ell)} \cup \left\{v\right\}, E^{(\ell+1)}, M^{(\ell+1)}\right\}$, which has a corresponding genotype matrix $B^{(\ell+1)} = \left[\begin{matrix}B^{(\ell)}, \va^{(\ell+1)};\vd^{(\ell+1)},1 \end{matrix}\right]$, where $\va^{(\ell+1)}$ is a length $\ell$ binary column vector indicating the ancestors of $v$ in $t^{(\ell+1)}$, $\vd^{(\ell+1)}$ is a length $\ell$ binary row vector indicating $v$'s descendants, and the submatrix $B^{(\ell)}$ is unchanged.

Each unique mutation tree is consistent with at most one binary genotype matrix $B$ that can be resolved in linear time\cite{gusfield_efficient_1991}, and vice versa. Consequently, we establish equivalent representations for binary genotype matrices as $t^{(1)} \equiv B^{(1)}$, $t^{(\ell)} \equiv B^{(\ell)}$, and $t^{(n)} \equiv B^{(n)}$. Without loss of generality, we will order the rows and columns of $B$, and its submatrices $B^{(\ell)}$, so that if the $v$-th row of $B^{(\ell)}$ represents the genotype of clone $v$, then the $v$-th column of $B^{(\ell)}$ corresponds to the clone defining mutation $M_v$ for clone $v$. As such $B^{(\ell)}_{v,v} = 1$ for any $v$. 
This row and column ordering guarantees that if $B^{(\ell)}_{v,w} = 1$, then either $v = w$ or clone $v$ is an ancestor of clone $w$ in $t^{(\ell)}$. 

\subsection{Motivating the approximate posterior}\label{sec:motivation-posterior}
\newcommand{\vb}{\mathbf{v}}

Orchard's goal is to sample without replacement from the mutation tree posterior defined by:
\begin{align}\label{eq:posterior}
    P(B|D) =  \frac{P(B,D)}{P(D)},
\end{align}
where $B$ is a genotype matrix and $D$ is a data set for $n$ point mutations across $m$ bulk samples. Note that computing $P(B,D) = \int_U P(B,U,D)dU$ requires marginalizing over the clonal proportion matrix $U$. According to Cayley's formula \cite{cayley_theorem_1889}, $(n+1)^{n-1}$ unique mutation trees can be constructed with $n$ mutations and a root node $r$. Consequently, if $D$ contains data for a large number of mutations, then it is intractable to compute the normalizing constant, $P(D) = \sum _B\int_U P(B,U,D)dUdB$, since this requires instantiating $(n+1)^{n-1}$ perfect-phylogeny-compatible $B$-matrices \cite{cayley_theorem_1889}, and integrating over the possible values of $U$ for each $B$. Fortunately, there are a variety of Markov Chain Monte Carlo (MCMC) techniques which do not require computing $P(D)$. These techniques rely on the following:
 
\begin{align}\label{eq:factorized-posterior-pre}
    P(B|D) 
    &\propto P(B,D) \nonumber\\ 
    &\propto \int_U P(B,U,D)dU \nonumber\\ 
    &\propto \int_U P(D|B,U)P(U|B)P(B)dU.
\end{align}
Eq \ref{eq:factorized-posterior-pre} can be used to design an MCMC sampling algorithm, where $B$-matrices are sampled and subsequently evaluated using the likelihood $P(D|B,U)$ by computing (or estimating) the integral over $U$, see, e.g.,  \cite{wintersinger_reconstructing_2022}. MCMC methods can be slow to converge and mix, and if $P(B|D)$ has low entropy, then many samples will need to be drawn before $k$ distinct ones are obtained. Here, we propose an approximation to the posterior, $Q^\pi(B|D)$, from which one can sample trees containing all $n$ mutations by progressively adding one mutation at a time to a partial tree structure. We will then show how Gumbel-Max tricks \cite{kool_stochastic_2019, kool_ancestral_2020, paulus_gradient_2021} can be used to efficiently sample without replacement from $Q^\pi(B|D)$.

We define $Q^\pi(B|D)$ by choosing a fixed order in which mutations are added to the tree; the accuracy of its approximation of $P(B|D)$ depends on this ordering. We denote a mutation order with the vector $\mathbf{\pi}$ representing a permutation, whose $\ell$-th element, ${\pi}_{\ell}$, is the index of the $\ell$-th mutation added. Each mutation appears in $\pi$ exactly once. We use the notation $D^{(\ell+1)}$ to represent the data associated with the first $\ell + 1$ mutations in $\mathbf{\pi}$. Given $\mathbf{\pi}$, we write:
\begin{align}
    Q^\pi(B|D) 
    &= \prod_{\ell=1}^{n-1} P(B^{(\ell+1)} | D^{(\ell+1)}, B^{(\ell)}) \label{eq:approximate-factorized-posterior}\\
    &\approx \prod_{\ell=1}^{n-1} P(B^{(\ell+1)} | D, B^{(\ell)}) = P(B|D)\label{eq:factorized-posterior}
\end{align}
Here, the genotype matrix $B^{(\ell+1)}$ is obtained from $B^{(\ell)}$ by appending $\va^{(\ell+1)}$, representing the ancestors of the new mutation in $t^{(\ell+1)}$, and $\vd^{(\ell+1)}$, representing its descendants. Eq \ref{eq:factorized-posterior} follows directly from factoring $P(B|D)$ into sets of binary random variables, i.e., $\{(\va^{(2)}, \vd^{(2)}), (\va^{(3)}, \vd^{(3)}), \dots, (\va^{(n)}, \vd^{(n)})\}$ associated with the elements in the newly added row and column in $B^{(\ell+1)}$ versus $B^{(\ell)}$ and noting that $B^{(1)}$ is a $1 \times 1$ binary matrix containing a $1$. See Appendix \ref{appendix:approximate-posterior-derivations} for more details. Sampling from $Q^{\pi}(B|D)$ in place of $P(B|D)$ is dramatically more efficient. Specifically, computing each factor $P(B^{(\ell+1)}|D, B^{(\ell)})$ requires evaluating all mutation trees of size $n$ that include $B^{(\ell+1)}$ as a submatrix. In contrast, computing $Q^{\pi}(B^{(\ell+1)}|D^{(\ell+1)})$ requires evaluating only mutation trees of size $\ell+1$ that contain $B^{(\ell)}$ as a submatrix. The difference in computational complexity is super-exponential.

In general, Eq \ref{eq:approximate-factorized-posterior} is equal to Eq \ref{eq:factorized-posterior} if and only if:
\begin{equation}\label{eq:conditional-independence-assumption}
    B^{(\ell+1)} \upmodels D \setminus D^{(\ell+1)} | B^{(\ell)}, D^{(\ell+1)},
\end{equation}
i.e., that, given $B^{(\ell)}$ and the data for the first $\ell+1$ mutations, $D^{(\ell+1)}$,  $B^{(\ell+1)}$ is conditionally independent from the data for the last $n - (\ell+1)$ mutations in $\pi$, $D \setminus D^{(\ell+1)}$. This conditional independence assumption states that data for mutations not yet in the tree can be ignored when placing new mutations. Ordering the mutations so that no mutation is placed in the tree before its ancestors will often ensure that this approximation is accurate. However, there are counterexamples where this is not true. We discuss this issue in more detail in Appendix \ref{appendix:mutation-ordering}.

Evaluating each placement of $\pi_{\ell+1}$ into $t^{(\ell)}$ requires computing the conditional likelihood $P(B^{(\ell+1)} | D^{(\ell+1)}, B^{(\ell)})$. This term can be rewritten as follows:

\begin{equation}\label{eq:add-mutation-posterior}
    P(B^{(\ell+1)} | D^{(\ell+1)}, B^{(\ell)}) \propto P(D^{(\ell+1)}|B^{(\ell+1)})P(B^{(\ell+1)}| B^{(\ell)}).
\end{equation}
We discuss computing the data likelihood $P(D^{(\ell+1)}|B^{(\ell+1)})$ in Section \ref{sec:complete-data-likelihood}. Eq \ref{eq:add-mutation-posterior} introduces a new term, $P(B^{(\ell+1)}|B^{(\ell)})$, which we define as:

\newcommand{\vp}{{\mathbf{p}}}
\newcommand{\qm}[1]{\textbf{QM: #1}}

\begin{equation}
    P(B^{(\ell+1)}|B^{(\ell)}) = 
    \begin{cases}
    \tau^{-1} & \begin{aligned}
       & \text{if } B^{(\ell+1)} \text{ is a perfect phylogeny genotype matrix and } \\ 
       & B^{(\ell+1)} = \left[ B^{(\ell)}, \va^{(\ell+1)} ; \vd^{(\ell+1)}, 1 \right] \\ 
    \end{aligned}\vspace{0.3cm} \\
    0 & \text{ otherwise},
    \end{cases}
\end{equation}
where $\tau$ is the number of possible settings of $\va^{(\ell+1)}$ and $\vd^{(\ell+1)}$ such that $B^{(\ell+1)}$ is a perfect phylogeny genotype matrix.

\subsection{Orchard algorithm}\label{sec:orchard-algorithm}

\begin{figure}[ht]
\includegraphics[scale=0.17]{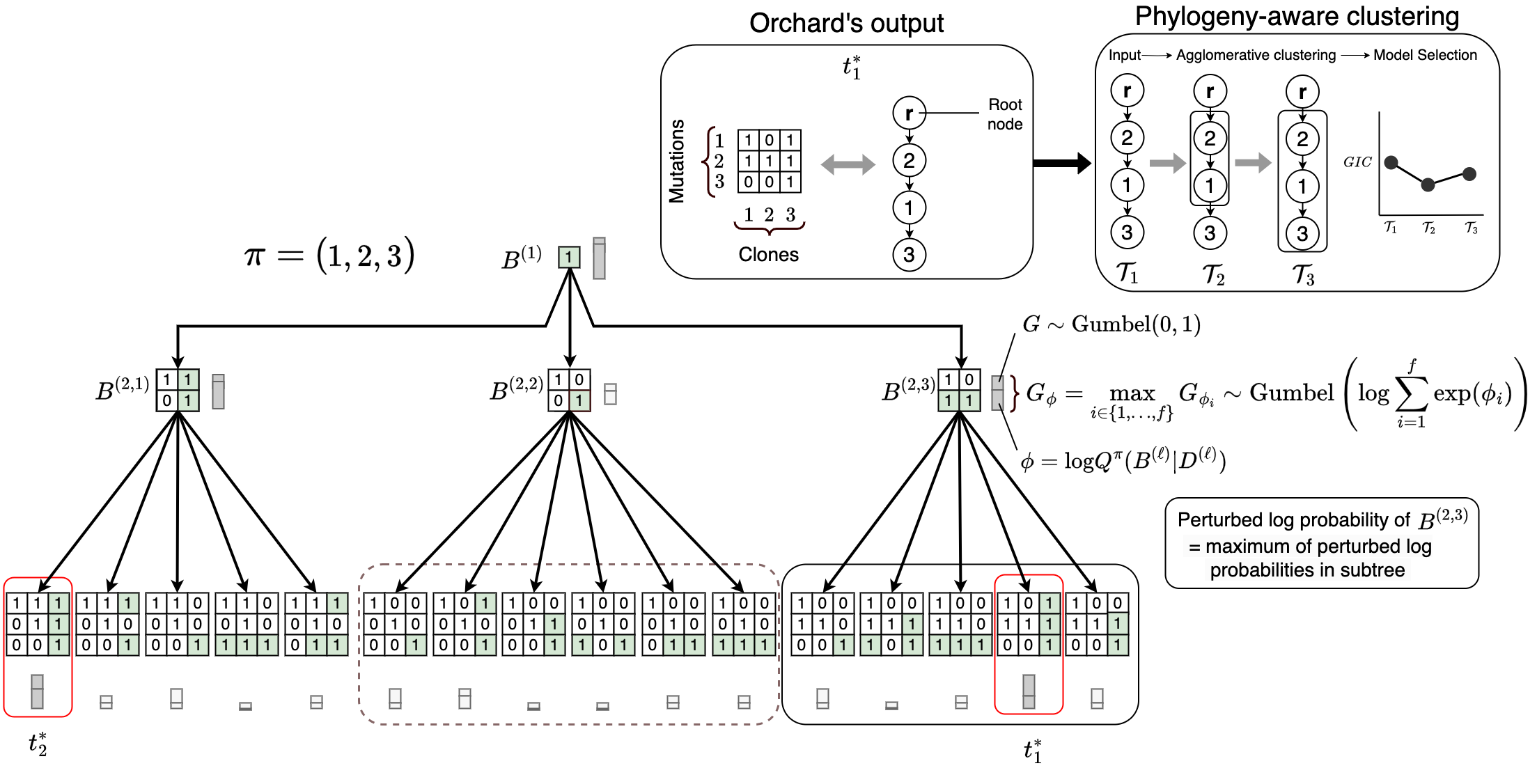}
\caption{Example of Orchard's mutation tree search with k=2, $f=\infty$. Mutation trees are depicted using genotype matrices. Search begins with a genotype matrix $B^{(1)}$ containing the first mutation in $\pi$. During each iteration, the best tree $t^{(\ell)}$ is popped from the queue and extended. The extensions are scored and reintroduced into the queue. Only the $k$ trees with the highest scores in the queue are kept, while others are discarded. The bars next to each genotype matrix indicate its perturbed log probability, $G_{\phi}$. Bars with grey fill correspond to the top-$k$ trees that are retained and extended. Genotype matrices within dashed boxes denote parts of the search space that are not explored further. Orchard's best reconstructed tree can then be input into the phylogeny-aware clustering algorithm. This algorithm conducts agglomerative clustering on the mutation trees to produce a set of clone trees. Each clone tree's set of clones is scored, and the algorithm yields the clone tree that minimizes the Generalized Information Criterion (GIC). See Section \ref{sec:pac} and Appendix \ref{appendix:phylogeny-aware-clustering} for complete details.} \label{fig:orchard-search}
\end{figure}

Orchard performs beam search in a search tree to generate $k$ samples without replacement from $Q^{\pi}(B|D)$. In this search tree, each node represents a unique mutation tree, leaf nodes are complete mutation trees, and internal nodes are partial mutation trees (see Figure \ref{fig:orchard-search}).
Each node in the search tree is labeled with the perturbed (unnormalized) log probability of its corresponding mutation tree. The search tree has a total of $(n+1)^{n-1}$ leaves, representing each unique mutation tree that can be constructed from $n$ mutations. Orchard's search routine finds the top-$k$ leaves of the search tree with the largest perturbed log probabilities without exhaustive exploration \cite{kool_ancestral_2020}. To further reduce run times, Orchard employs heuristic optimizations that, empirically, do not compromise its accuracy.


\subsubsection{Background on Ancestral Gumbel-Top-k trick}
Sampling by finding the largest perturbed log probabilities is an example of a \textit{Gumbel-Max trick}.
These tricks derive from the motivating observation that if $\phi_i$ is the unnormalized log probability of the $i$-th item in a set, $\left\{\phi_1, \dots, \phi_f\right\}$, and $G^{(i)} \sim \text{Gumbel}(0,1)$ is an i.i.d. sample from the standard Gumbel distribution -- where $\text{Gumbel}(\mu, \beta)$ has a location parameter $\mu$ and a scale parameter $\beta$ -- then the following relationship holds:
\begin{equation}\label{eq:gumbel-max-trick}
I = \argmax_i\left\{\phi_{i} + G^{(i)}\right\} \sim \text{Categorical}\left(\frac{\text{exp}(\phi_{i})}{\displaystyle\sum_{j = 1}^f\text{exp}(\phi_{j})}, i \in \{1,2,\dots,f\}\right).
\end{equation}
Eq \ref{eq:gumbel-max-trick} is recognized as the original Gumbel-Max trick \cite{maddison_ast_2014}, and we refer to the quantities $G_{\phi_i} = \phi_{i} + G^{(i)}$ as \textit{perturbed log probabilities}, noting that $G_{\phi_i} \sim \text{Gumbel}(\phi_i, 1)$. It shows that finding the index of the largest perturbed log probability in a set is equivalent to sampling from a categorical distribution, defined by unnormalized log probabilities associated with each element. This powerful observation can be used to convert various combinatorial optimization algorithms into sampling algorithms. In a simple example, one can draw $k$ samples without replacement from $\left\{\phi_1,\dots,\phi_f\right\}$, using an extension of the Gumbel-Max trick called the \textit{Gumbel-Top-k} trick\cite{paulus_gradient_2021}, by finding the $k$ largest perturbed log probabilities in the set $\{G_{\phi_1}, G_{\phi_2}, \dots, G_{\phi_f}\}$ by, e.g., sorting the set\cite{paulus_gradient_2021}.

Applying Gumbel-Top-k to sample from either $P(B|D)$ or $Q^{\pi}(B|D)$ would require instantiating all $(n+1)^{n-1}$ unique trees over $n$ mutations. Fortunately, the conditional independence structure of $Q^\pi(B|D)$ supports a tractable extension called the \textit{Ancestral Gumbel-Top-k} trick. The Ancestral Gumbel-Top-k trick can be used alongside a branch-and-bound search to find the highest-scoring trees, leveraging a property of Gumbel variables known as \textit{max-stability}. Specifically, the maximum of a set of independent Gumbel random variables $\{G_{\phi_1}, G_{\phi_2}, \dots, G_{\phi_f}\}$ follows a Gumbel distribution with a location equal to the log-sum-exp of their log probabilities $\{\phi_1, \phi_2, \dots, \phi_f\}$ \cite{gumbel_statistical_1954, kool_ancestral_2020, maddison_ast_2014}, i.e.,
\begin{equation}
    \max_{i \in \{1,2,\dots,f\}}G_{\phi_i} \sim \text{Gumbel}\left(\text{log}\sum_{i=1}^f\text{exp}(\phi_i),1\right)\label{eq:max-stability}.
\end{equation}
Max-stability can be used to determine the maximum perturbed log probability in a subtree of the search tree. This property allows a branch-and-bound beam search to efficiently find the top $k$ largest perturbed log probabilities in the search tree, avoiding exhaustive exploration. To illustrate this, first, let $B^{(\ell)}$ be a genotype matrix with log probability $\phi = \log Q^{\pi}(B^{(\ell)}|D^{(\ell)})$, and let $B^{(\ell+1,i)}$ be its $i$-th extension (out of $f$ possible ones) with log probability $\phi_i = \log Q^{\pi}(B^{(\ell+1,i)}|D^{(\ell+1)})$. Note that $\phi$ is the log-sum-exp of the set $\{\phi_i\}_{i=1}^f$:
\begin{align}
\exp(\phi) &= Q^{\pi}(B^{(\ell)} | D^{(\ell)}) = Q^{\pi}(B^{(\ell)} | D^{(\ell+1)}) \\
&= \sum_{i=1}^f P(B^{(\ell+1,i)}|B^{(\ell)},D^{(\ell+1)})Q^{\pi}(B^{(\ell)} | D^{(\ell)})\label{eq:gumbel-location-marg} \\
&=\sum_{i=1}^f\text{exp}(\phi_i)\label{eq:gumbel-location-sum-exp},
\end{align}
where Eq \ref{eq:gumbel-location-marg}, \ref{eq:gumbel-location-sum-exp} result from marginalizing over all $f$ extensions of $B^{(\ell)}$.
Thus we can write
\begin{equation}\label{eq:max-of-subtree}
    \max_{i \in \{1,2,\dots,f\}}G_{\phi_i} = G_{\phi} \sim \text{Gumbel}(\text{exp}(\phi)).
\end{equation}
So, if computing $\phi$ is tractable, then one can determine the maximum perturbed log probability of the set ${\phi_i}_{i=1}^f$ without computing any of the $\phi_i$s. Additionally, because Eq \ref{eq:max-of-subtree} can be applied recursively to each extension of $B^{(\ell)}$, and so on, down to the leaves of the search tree rooted at $B^{(\ell)}$, the perturbed log probability of $B^{(\ell)}$ is also the maximum of any complete mutation trees that have $B^{(\ell)}$ as a submatrix.

Orchard leverages this property by using $G_\phi$ as an estimate of the best complete tree that can be found by sampling the leaves in the subtree below $B^{(\ell)}$ and employs a beam search to explore only those parts of the search tree containing the top-$k$ leaves.

\subsubsection{Computing the ``shifted" perturbed log probabilities of extensions}\label{sec:shifting-perturbed-log-probabilities}

When Orchard extends $t^{(\ell)}$, it has already sampled $G_{\phi}$. For each extensions $t^{(\ell+1,i)}$ that's considered, $G_{\phi_i}$ is sampled. However, each $G_{\phi_i}$ must be sampled conditionally such that $G_{\phi} = \max_{i \in \{1,2,\dots,f\}}G_{\phi_i}$ (Eq \ref{eq:max-of-subtree}). Orchard achieves this by initially sampling the set  $\{G_{\phi_1}, G_{\phi_2}, \dots, G_{\phi_f}\}$, then applying a transformation, as described in \cite{kool_ancestral_2020}, to ensure this condition is satisfied, resulting in a new set $\{\bar{G}_{\phi_1}, \bar{G}_{\phi_2}, \dots, \bar{G}_{\phi_f}\}$. We briefly describe this transformation here, but encourage the reader to refer to \cite{kool_ancestral_2020} for a complete proof of correctness. Let $Z = \max_j G_{\phi_j}$ and $T = \bar{G}_{\phi}$. We assume $\bar{G}_{\phi}$ is also readily available, as it can be initialized to $0$ at the start of the algorithm. The transformation sets $Z = T$, and shifts each $\{G_{\phi_1}, G_{\phi_2}, \dots, G_{\phi_f}\}$ to be less than or equal to $Z$ using the truncated Gumbel CDF: 

\begin{align}\label{eq:gumbel-transform}
    \bar{G}_{\phi_i} &= F^{-1}_{\phi_i,T}(F_{\phi_i, Z}(G_{\phi_i})) \\
    &= \text{log}(\text{exp}(-T)) - \text{exp}(-Z) + \text{exp}(-G_{\phi_i})\nonumber,
\end{align}
where $\bar{G}_{\phi_i}$ is the ``shifted" perturbed log probability of the $i$-th extension, $F_{\phi,T}(g)$ is the CDF of the Gumbel distribution truncated at $T$, and $F^{-1}_{\phi,T}(u)$ is the inverse CDF of the Gumbel distribution truncated at $T$. Please see \cite{kool_ancestral_2020} for a numerically stable version of Eq \ref{eq:gumbel-transform}. 

\subsubsection{Stochastic beam search}

Algorithm \ref{alg:orchard} contains the pseudocode for the Orchard algorithm. Orchard's search routine can be considered a \textit{stochastic beam search}, and is adapted from the algorithms described in \cite{kool_stochastic_2019, kool_ancestral_2020}. Orchard keeps track of partial trees using a queue with a fixed size $k$, a user-defined parameter called the \textit{beam width}. At each iteration, the partial tree $t^{(\ell)}$ with the largest $\bar{G}_{\phi}$ is popped from the queue and extended. For each extension $t^{(\ell+1,i)}$ that we consider, we compute a perturbed log probability, but to do so, we need to compute, or estimate, $P(D^{(\ell+1)} | B^{(\ell+1,i)})$, which is a computationally intensive operation.
As such, we use a heuristic, denoted by the function $H(t^{(\ell)},f)$, which uses an analytical approximation to score all extensions based on their adherence to the ISA, and selects the top-$f$ extensions based on these scores. The parameter $f$ is user-defined, and is called the \textit{branching factor}. In practice, we find this heuristic decreases Orchard's run time without compromising its accuracy. Complete details for this heuristic are provided in Appendix \ref{appendix:placing-mutations}.

\SetKwInput{Parameter}{parameters}
\SetKwInput{Input}{input}
\SetKwInOut{Output}{output}
\begin{algorithm}[h]
\SetAlgoLined
\caption{Orchard}\label{alg:orchard}
\Input{A dataset $D$ containing allele frequency data and copy number states for $n$ mutations in $m$ tissue samples}
\Parameter{k (\textit{beam width}), f (\textit{branching factor}), $\pi$ (\textit{mutation order})}
\Output{$t^*_1, t^*_2, \dots, t^*_k \text{, the top-}k \text{ trees} $}
\BlankLine
\tcp{Initialize first tree assuming that $\pi = \left[\begin{matrix}1 & 2 & \dots & n\end{matrix}\right]$}
$t^{(1)} = \left\{V^{(1)} = \{r, 1\}, E^{(1)}=\left\{(r,1)\right\}, M^{(1)}=\left[\begin{matrix} 1 & -1 & \dots & -1 \end{matrix}\right]\right\}$ \\
\BlankLine
$\bar{G}_{\phi} = 0, G_{\phi} = 0$ \tcp{initialize scores} 
$Q = \left[(t^{(1)}, \bar{G}_{\phi}, G_{\phi})\right]$ \\

\While{$Q \text{ is not empty}$}{
 $(t^{(\ell)}, \bar{G}_{\phi}, G_{\phi}) = \text{ take and remove the first element of }Q$ \\
 \BlankLine
 \tcp{$\text{H}(t^{(\ell)},f)$ returns a set of $f$ indices corresponding to extensions of $t^{(\ell)}$ with the best adherence to the ISA -- this helps speed up lines 14, 16}
 $S = \left\{G_{\phi_i} = \phi_i + G^{(i)} : i \in \text{H}(t^{(\ell)},f)\right\}\label{alg:orchard:branching-factor}$  \\
  \eIf{$\ell + 1 == n$}{
 $k = k - 1$ \\
 $\textbf{yield} \hspace{.2cm} t^* = 
 \argmax_iS$ \tcp{$\text{yield best tree in } S$} \BlankLine
 $\text{Return to line } 5$
}   
{
\tcp{"Shift" perturbed log probabilities using Eq  \ref{eq:gumbel-transform}}
$Z = \max_iS$ \\
 $T = G_{\phi}$ \\
 $\bar{S} = \left\{\bar{G}_{\phi_i} = \text{log}\left(\text{exp}(-T) - \text{exp}(-Z) + \text{exp}(-G_{\phi_i})\right) : G_{\phi_i} \in S \right\}$ \\

$\text{add each } (t^{(\ell+1)}, \bar{G}_{\phi_i},G_{\phi_i}) \text{ to } Q$

 $\text{Sort } Q \text{ in descending order according to } \bar{G}_{\phi} \text{, discard all but top-$k$}$ 
 }
}
\end{algorithm}

\subsubsection{Approximating the MLE of the clonal proportion matrix}\label{appendix:computing-clonal-proportion}

Ideally, if we seek to score a mutation tree represented by its binary genotype matrix $B$, we would integrate out the clonal proportions $U$. In this case, the data likelihood for $B$, $P(D|B)$, assuming a uniform prior on $U$, is given by 

\begin{equation}
    P(D | B) \propto \prod_s \int^{1 \geq \sum_v U_{vs}}_{\mathbf{0}}  P(D_s|B,U_{:s}) dU_{:s},
\end{equation}
where $D_s$ are the data associated with sample $s$ and the integration is over non-negative values of $U_{vs}$ such that $\sum_v{U_{vs}} \leq 1$. 
We know of no analytical solution to this integral, instead we use a point estimate of $U$, $U^*$ and use the approximation
\[
P(D | B) \approx \prod_s P(D_s|B,U^*_{:s}).
\]
A good point estimate of $U$ would be its maximum likelihood estimate (MLE) given $D$ and $B$. Instead, to speed up Orchard, we use the projection algorithm \cite{jia_efficient_2018, ray_exact_2019} which finds a point estimate $U^*$ by optimizing the following quadratic approximation to the binomial likelihood of $U$ given $B$ and $D$: 

\begin{equation}\label{eq:projection-optimization}
LP(U^*; B, \widehat{F}, W) = \min_{F^*,U^*} \left\lVert W\odot\left(\widehat{F} - F^*\right)\right\rVert^2 \text{ s.t. } \mathbf{1}^TU^* \leq \mathbf{1}, U^* \geq 0, F^* = BU^*,
\end{equation}
where $\lVert \cdot \rVert$ is the Frobenius norm, $\mathbf{1}$ is a vector of ``1"s, $\widehat{F}$ are the observed mutation frequencies, $W$ is an $n \times m$ matrix of inverse-variances for each mutation in each sample derived from $D$, and $\odot$ is the Hadamard, i.e., element-wise product. We use the same definition for $W$ as described in \cite{ray_exact_2019, wintersinger_reconstructing_2022}.



\subsubsection{Approximating the complete data likelihood}\label{sec:complete-data-likelihood}

Orchard evaluates a genotype matrix $B$ and its corresponding clonal proportion matrix $U$ using an approximation of the complete data likelihood of the input data $D$. This approximation assumes that the observed bulk read count data for each locus is generated according to a binomial sampling model, i.e., 
\begin{equation}
    b_{js} \sim \text{Binom}(N_{js},\lambda_{js}),
\end{equation}
where $b_{js}$ is the count of reads containing the variant allele $j$ in sample $s$, $N_{js} = a_{js} + b_{js}$ is the total number of reads mapped to the locus containing the variant allele $j$ in sample $s$, and $\lambda_{js}$ is the variant allele frequency for the variant allele $j$ in sample $s$. We do not observe $\lambda_{js}$, but instead infer it from the mutation frequency matrix $F$. Given a genotoype matrix $B$ and a clonal proportion matrix $U$, we calculate the mutation frequency matrix with $F = BU$ (Eq \ref{eq:F-equals-BU}). We can use the following formula to estimate the variant allele frequency that generated the observed bulk read count data for mutation $j$ in sample $s$:

\begin{equation}
    \lambda_{js} = F_{js}\omega_{js},
\end{equation}
where $\lambda_{js}$ is the inferred variant allele frequency, $F_{js}$ is the frequency of mutation $j$ in sample $s$, and $\omega_{js}$ is the variant read probability provided as an input along with the read count data. We then compute the approximate complete data likelihood as follows:

\begin{align}
    P(D|B,U) &\approx \prod_s{L(s, F_{:s})} \label{eq:approximate-data-likelihood}\\ 
L(s,F_{:s}) &= \prod_v{\text{Binom}(b_{vs} | a_{vs}+b_{vs}, \omega_{vs} F_{vs})}.\label{eq:approximate-data-likelihood-factor}
\end{align}
We assume that the samples are exchangeable in Eq \ref{eq:approximate-data-likelihood}. To complete our definition for Eq \ref{eq:approximate-data-likelihood-factor}, we define $\text{Binom}(b | N, \lambda) = \binom{N}{b}\lambda^b(1-\lambda)^{a}$ as the binomial probability of observing $b$ variant reads out of a total of $N = a + b$ reads mapping to its locus when the variant allele frequency is $\lambda$.

\subsection{Phylogeny-aware clustering}\label{sec:pac}
Orchard contains a ``phylogeny-aware" clustering algorithm that uses a reconstructed mutation tree and its corresponding mutation frequency matrix $F$ to cluster mutations into subclones and resolve clone trees. This algorithm performs agglomerative clustering and joins adjacent nodes $u$ and $v$ in a tree using the following formula: 

\begin{align*}
    \text{min}_{u,v}{d(u,v)}, \\ 
    d(u,v) &= \frac{n_un_v}{n_u + n_v}\lVert \bar{F}_u - \bar{F}_v \rVert^2,
\end{align*} 
 where $n_u$ is the number of mutations associated with node $u$, $\bar{F}_u$ is the average mutation frequency of the mutations contained in node $u$, and $\lVert \cdot \rVert$ is the Euclidean norm. This definition for $d(u,v)$ is known as \textit{Ward's method}. The agglomerative clustering is repeated, successively joining adjacent nodes, until it reaches a clone tree consisting of only the root node and a single node containing all mutations. 
Note that only parent-child pairs, where the child is the parent's only child, are considered for joining. Once no such pairs remain, any adjacent nodes can then be joined. The algorithm outputs $n$ different clusterings containing $1,\dots,n$ clones. Each clone tree containing $c$ clones has a corresponding $n \times c$ binary matrix $Z$, where $Z_{ji} = 1$ if and only if mutation $j$ originated in clone $i$. Associated with each clone $i$ in each sample $s$ is its variant allele frequency, $\lambda_{is}$. If  mutation $j$ is assigned to clone $i$, then it is assumed that the observed variant allele frequency for mutation $j$ in sample $s$, $\widehat{\lambda}_{js}$, is a noisy observation of $\lambda_{is}$, such that $\widehat{\lambda}_{js} = \lambda_{is} + \epsilon$, where $\epsilon$ is a noise term. So, given $F$, $Z$, and $D$, the likelihood of a clustering can be expressed as:

\begin{equation}\label{eq:pac-likelihood}
 L(F|D, Z) = \sum_{i=1}^c\sum_{j=1}^n\mathbb{I}(Z_{ji} = 1)\sum_{s=1}^m\text{log}\left(\text{Binom}(b_{js}|N_{js},\lambda_{is})\right),
\end{equation}
where $\text{Binom}(b|N,\lambda) = \binom{N}{b}\lambda^b(1-\lambda)^a$ is defined as the binomial probability of observing $b$ variant reads out of a total of $N = a + b$ reads mapping to its locus when the variant allele frequency is $\lambda$, and $\mathbb{I}(Z_{ji} = 1)$ is the indicator function which evaluates to 1 if mutation $j$ is assigned to clone $i$, and 0 otherwise. Maximizing Eq \ref{eq:pac-likelihood} may result in overfitting, where each mutation is assigned to its own clone. To mitigate this issue, the clone tree with the clustering that minimizes the Generalized Information Criterion (GIC) \cite{kim_consistent_nodate} is yielded. The implementation details of this algorithm are provided in Appendix \ref{appendix:phylogeny-aware-clustering}.

\section*{Results}
\subsection{Evaluation Overview}

We compare Orchard with two state-of-the-art MSPP reconstruction algorithms: CALDER \cite{myers_calder_2019}, and Pairtree \cite{wintersinger_reconstructing_2022}. For CALDER we used its non-longitudinal model and v8.0.1 of the Gurobi optimizer. CALDER only outputs a single tree which is optimal under its mixed integer linear programming formulation, or otherwise fails to output a valid tree. Pairtree uses Markov Chain Monte Carlo to sample trees from a data-implied posterior over trees. We set Pairtree's \textit{trees-per-chain} parameter to $5000$, and otherwise used the default parameters. We ran the Pairtree algorithm in parallel on 32 CPU cores, generating $160000$ trees ($32 \times 5000$). All Orchard experiments used the same branching factor $f=20$, but we evaluated two beam widths: $k=1$, and $k=10$. We also ran the Orchard algorithm in parallel on 32 CPU cores; generating $32$ trees ($32 \times 1$) and $320$ trees ($32 \times 10$), respectively. Parallel instances of Orchard sample trees independently, so we are only guaranteed a minimum of $1$ or $10$ unique trees, respectively. 
Running parallel instances of Orchard and Pairtree means that each algorithm was effectively executed 32 times on each dataset, with each parallel instance having a unique random seed. CALDER, being deterministic when run on identical computing hardware with the same inputs, requires only a single run to obtain its best reconstruction. 

We assess the mutation frequency matrix $F$ and genotype matrix $B$ for the tree(s) output by a MSSP reconstruction method by comparing them with a baseline mutation frequency matrix $F^{(\text{baseline})}$ and, when available, a set of ground-truth genotype matrices. For simulated bulk data, $F^{(\text{baseline})}$ corresponds to the ground-truth mutation frequency matrix $F^{(\text{true})}$ used to generate the simulated VAF data. For real bulk data where an expert-defined clone tree is available, $F^{(\text{baseline})}$ corresponds to a \textit{maximum a posteriori} (MAP) mutation frequency matrix $F^{(\text{MAP})}$ that we fit to the expert-defined tree (see Appendix \ref{appendix:rprop}). Each mutation in $F^{(\text{MAP})}$ is assigned a frequency according to the MAP mutation frequency of the subclone the mutation belongs to in the expert-defined clone tree. We compare the reconstructed mutation frequencies in $F$ to $F^{(\text{baseline})}$ using a metric called the \textit{log perplexity ratio} \cite{wintersinger_reconstructing_2022}. The log perplexity ratio is the log of the ratio between the perplexity of $F$ and the perplexity of $F^{(\text{baseline})}$. A negative log perplexity ratio indicates that the mutation frequency matrix $F$ for the tree(s) reconstructed by a method agree with the VAF data better than $F^{(\text{baseline})}$. Perplexity is evaluated under a binomial probability model. We compare the reconstructed genotype matrix $B$ to the set of ground-truth genotype matrices using a metric called the \textit{relationship reconstruction loss} \cite{wintersinger_reconstructing_2022}. The relationship reconstruction loss compares the evolutionary relationships in $B$ to those in the set of ground-truth genotype matrices using the Jensen-Shannon divergence. The log perplexity ratio and relationship reconstruction loss are computed as likelihood-weighted averages over all distinct trees returned by a method. This approach improves the robustness and reliability of our evaluations by incorporating the uncertainty in the reconstructions. We also evaluate MSSP reconstruction methods based on their \textit{wall clock runtime}, representing the time (in seconds) each method requires to complete on a data set. For more details on these metrics see Appendix \ref{appendix:evaluation-metrics}.

\subsection{Orchard produces better reconstructions than competing methods on 90 simulated cancers}\label{sec:simulated-results}

We used the Pearsim software (https://github.com/morrislab/pearsim) from \cite{wintersinger_reconstructing_2022} to generate 90 simulated mutation trees and corresponding VAF data which adheres to the ISA.  The data sets vary in the number of mutations (10, 30, 50, 100, 150, 200), cancer samples (10, 20, 30, 50, 100), and sequencing depths (50x, 200x, 1000x). These simulations all adhere to the ISA.


\begin{table}[!ht]
\centering
\vspace{-0.35cm}
\begin{tabular}{|*{13}{c|}}  
\hline
\multirow{3}{4em}{} & \multicolumn{12}{|c|}{Simulated data set size (mutations)} \\ \cline{2-13}
 & \multicolumn{2}{|c}{10} & \multicolumn{2}{|c}{30} & \multicolumn{2}{|c}{50} & \multicolumn{2}{|c}{100} & \multicolumn{2}{|c}{150} & \multicolumn{2}{|c|}{200} \\ \cline{2-13}
& \textit{P} & \textit{R} & \textit{P} & \textit{R} & \textit{P} & \textit{R} & \textit{P} & \textit{R} & \textit{P} & \textit{R} & \textit{P} & \textit{R} \\ \hline
CALDER & 0 & 1 & 0 & 0 & 0 & 0 & 0 & 0 & 0 & 0 & 0 & 0 \\ \hline
Pairtree & 1 & 0 & 4 & 3 & 0 & 0 & 0 & 0 & 0 & 0 & 0 & 0 \\ \hline
Orchard (k=1) & \textbf{8} & \textbf{13} & \textbf{8} & \textbf{9} & \textbf{10} & 5 & 4 & 7 & 1 & \textbf{8} & 4 & 7 \\ \hline
Orchard (k=10) & 6 & 1 & 3 & 3 & 5 & \textbf{10} & \textbf{11} & \textbf{8} & \textbf{14} & 7 & \textbf{11} & \textbf{8} \\ \hline
\end{tabular}
\caption{Count of simulated mutation tree data sets, out of 15 per column, where a model had the best log perplexity ratio (P) or relationship reconstruction loss (R). Bold indicates column max.}
\label{table:simulated-results}
\end{table}

\begin{figure}
\includegraphics[scale=0.094]{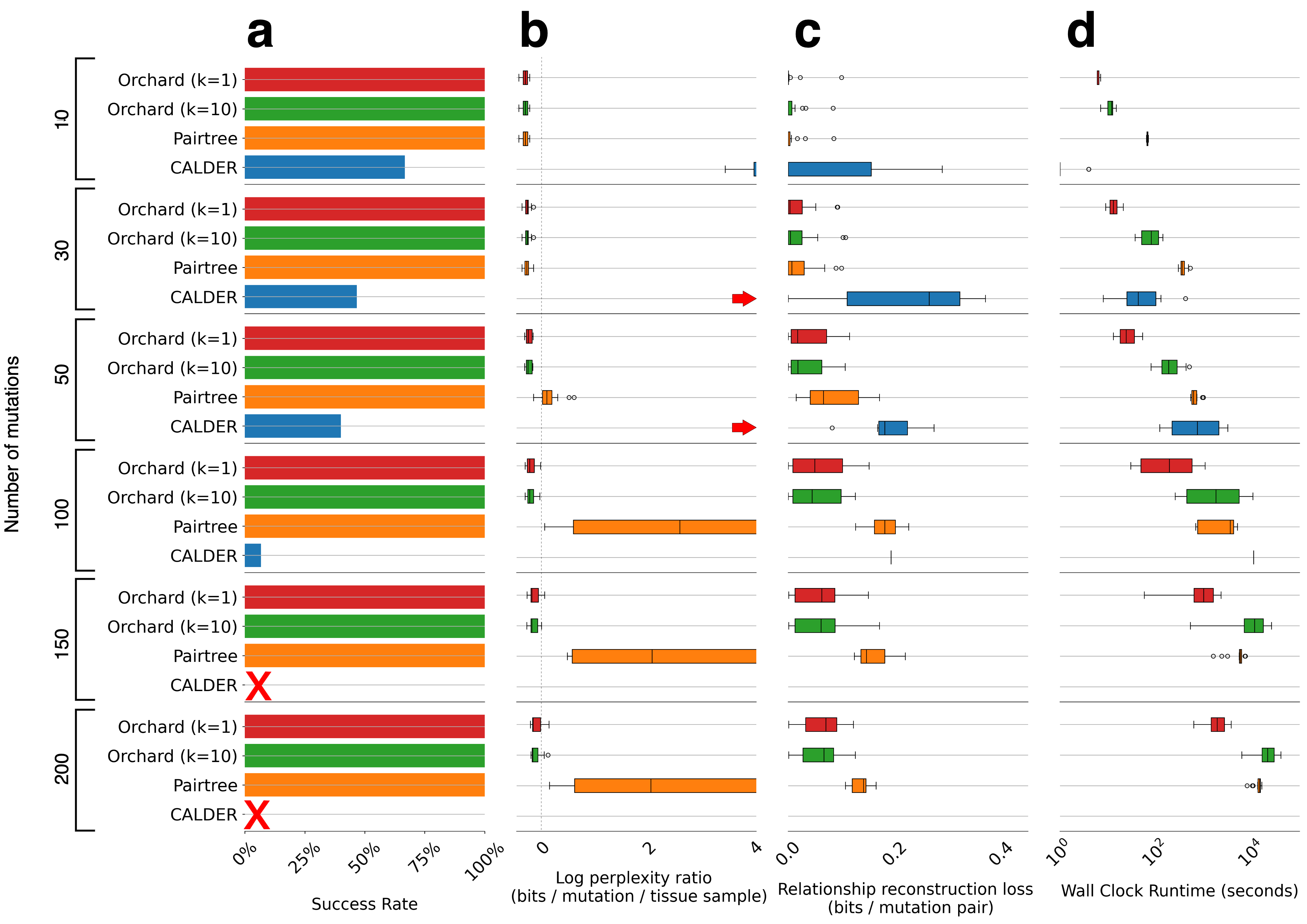}
\caption{Evaluation of reconstructions for 90 simulated mutation trees. Results are grouped by the size of the simulated mutation trees (rows), i.e., the \textit{problem size}. \textbf{a.} Bar plots show the percentage of data sets where a method produces at least one valid tree. A red $x$ means the method did not succeed on any of the data sets for that problem size. A red arrow means the results for the method on a problem size occur beyond the x-axis limit. The distributions, represented by box plots, in (b,c,d) only include data sets where the method was successful. \textbf{b.} The distribution of log perplexity ratios, a measure of VAF data fit. Ratios are relative to the perplexity of the ground truth mutation frequency matrix $F^{(\text{true})}$, and can be negative. Lower log perplexity ratios indicate better reconstructions. \textbf{c.} The distributions of relationship reconstruction loss for each method on a problem size. This loss can range between zero bits (complete match of pairwise relationships) and one bit (complete mismatch of pairwise relationships). \textbf{d.} The distributions of wall clock run time.} \label{fig:90mutsim_compare}
\end{figure}

We ran Orchard, CALDER, and Pairtree on each simulated data set a single time, and the box plots in Figure \ref{fig:90mutsim_compare}b-d show the distribution of results for each evaluation metric. Table \ref{table:simulated-results} contains counts of data sets, separated by problem size, where a method had the best log perplexity ratio (P) or relationship reconstruction loss (R). CALDER was the only method to fail on a reconstruction problem, its failure rate increased as the problem size increased, and it failed to produce a valid tree on any reconstruction problem with more than 100 mutations.

Figure \ref{fig:90mutsim_compare} and Table \ref{table:simulated-results} illustrate a problem-size-dependency in relative performance. Orchard generally outperformed Pairtree and CALDER on the small reconstruction problems (10-30 mutations) on all metrics. Although Orchard's reconstructions are only slightly better than Pairtree for these problems, Orchard is 5-10x faster. In contrast, Orchard consistently outperformed Pairtree and CALDER on all problems with 50 or more mutations. For these problems, in either setting of $k$, Orchard finds trees that are significantly better than those found by Pairtree or CALDER. Unfortunately, the run time required by Orchard increases significantly on larger problem sizes. Detailed run time breakdowns for each method across problem sizes can be found in Table \ref{suptable:simulated-runtimes}. Orchard ($k=1$) generally outperforms Orchard ($k=10$) on smaller problems (10-30 mutations), but this switches on larger problems. To understand this change, we inspected the trees recovered by Orchard ($k=1$ and $k=10$). In nearly all 10 or 30-mutation tree reconstruction problems, both Orchard configurations found the same best tree. However, Orchard ($k=10$) yields nine other sub-optimal trees which are included in the weighted averages used to compute the metrics resulting in slightly worse average scores. On larger problems, the trees reconstructed by Orchard ($k=10$) are generally much better than those reconstructed by Orchard ($k=1$).

\subsection{Orchard reconstructs more plausible trees on 14 B-progenitor acute lymphoblastic leukemias}\label{sec:real-results}

We applied Orchard, CALDER and Pairtree on real bulk data from 14 B-progenitor acute lymphoblastic leukemias (B-ALLs) originally studied in \cite{dobson_relapse-fated_2020}. The B-ALL data sets vary in the number of subclones (between 5 and 17), represented by expert-defined subclones, tissue samples (between 13 and 90), and mutations (between 16 and 292). All of the B-ALL data sets had an approximate sequencing depth of 212x \cite{dobson_relapse-fated_2020}. Each B-ALL patient had samples taken at diagnosis and relapse, and then each of these samples was transplanted into immunodeficient mice resulting in multiple patient derived xenografts (PDXs). The VAF data for each B-ALL was used by experts to manually cluster mutations into subclones and construct clone trees, both of which were reviewed for biological plausibility \cite{dobson_relapse-fated_2020}.

\begin{figure}
\includegraphics[scale=0.22]{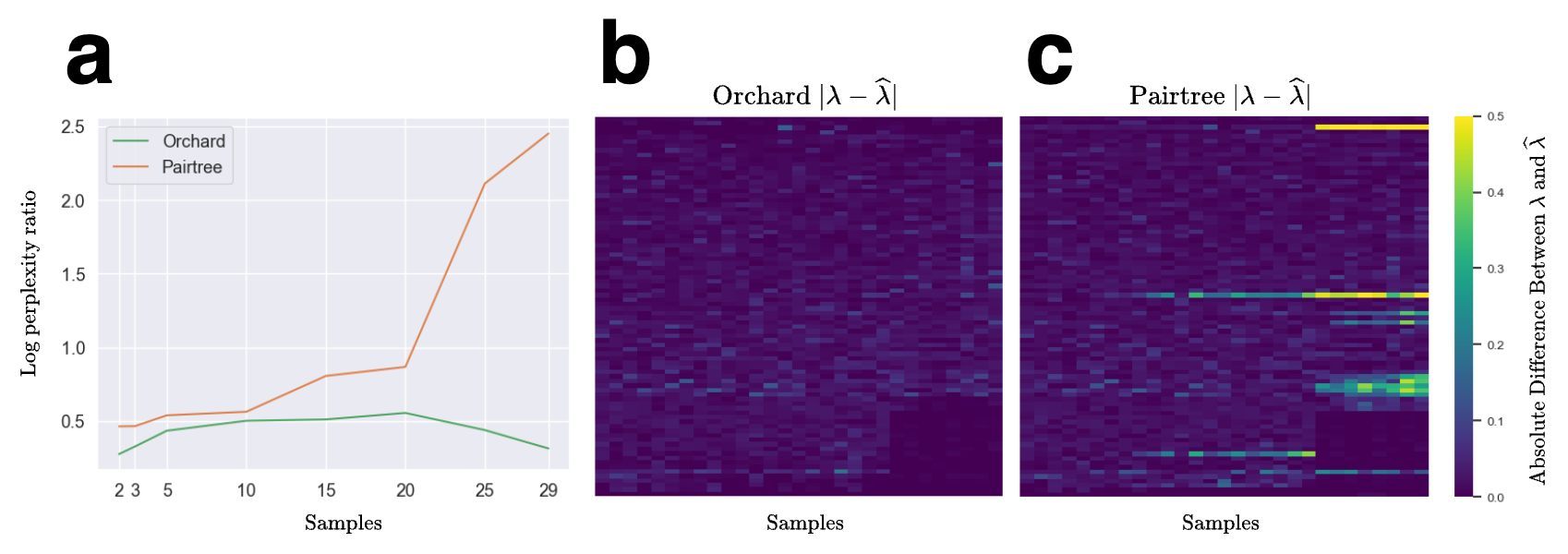}
\caption{Evaluation of reconstructions by Orchard and Pairtree for SJBALL022611. \textbf{a. } Log perplexity ratio for the trees reconstructed by Orchard and Pairtree as a function of the number of samples. Orchard's reconstructions are accurate regardless of the number of samples provided, while Pairtree's reconstructions worsen with more samples. \textbf{b,c. } Absolute difference between the VAFs inferred  by Orchard and Pairtree and the VAFs implied by the bulk data. Large values indicate divergence between VAFs inferred by a method and the VAFs implied by the data. VAFs inferred by Orchard adhere very closely to the data, while also adhering to the ISA. Pairtree's poorly reconstructed tree results in innaccurate VAF estimates for many mutations. The same row in each heatmap corresponds to same unique mutation, and each column corresponds to the same unique sample.}
\label{fig:ball-panel}
\end{figure}

Table \ref{table:ball-clone-tree-results} presents a comprehensive comparison of each method's log perplexity ratios on the 14 B-ALL clone tree reconstruction problems. Orchard generally outperformed Pairtree, achieving a slightly lower log perplexity ratio on 10/14 data sets, with an average reduction of -2.17e-5 bits. CALDER, on the other hand, failed on 2 of the data sets and performed the worst on those it did succeed on. It is unsurprising that both Orchard and Pairtree perform well on the B-ALL data, given that each data set contains at least 13 samples, and at most 17 clones. Typically, as the number of samples increases, the number of plausible trees typically decreases, making accurate reconstruction more feasible \cite{wintersinger_reconstructing_2022}.

In most practical scenarios, expert-defined subclones are unavailable. To emulate these conditions using the B-ALL data, we evaluate two approaches:

\begin{enumerate}
    \item The traditional approach of clustering mutations into subclones using VAF-based clustering algorithms and then constructing a clone tree from these subclones.
    \item Our new approach of reconstructing a mutation tree using the VAF data and subsequently utilizing this mutation tree to guide the inference of subclones.
\end{enumerate}
We provide a detailed assessment of both approaches in Appendix \ref{appendix:tree-based-analysis}. To assess the second approach, we first reconstructed mutation trees for each of the 14 B-ALL data sets using both Orchard (k=10) and Pairtree. We provide a comprehensive analysis of the mutation trees reconstructed by each method in Appendix \ref{appendix:b-all-mutation-trees}. Orchard and Pairtree reconstruct similarly accurate mutation trees on B-ALL data sets containing fewer than 50 mutations, but Orchard reconstructs significantly more plausible trees than Pairtree for B-ALLs with more than 50 mutations. 

The accuracy of Orchard's mutation tree reconstructions, compared to Pairtree, is highlighted by further analyzing the B-ALL data set SJBALL022611. This data set has 84 mutations identified across 29 different samples. First, Orchard (k=10) and Pairtree were used to reconstruct trees for this data using a varied number of samples (2,3,5,10,15,20,25,29). Figure \ref{fig:ball-panel}a shows the log perplexity ratio for the trees reconstructed by Pairtree and Orchard when varying the number of samples used. The baseline mutation frequencies were chosen to be the MAP mutation frequencies, $F^{(\text{MAP})}$, fit to the expert-defined clone tree for SJBALL022611. Counter intuitively, Figure \ref{fig:ball-panel}a demonstrates that as more samples are used during reconstruction, the mutation frequencies reconstructed by Pairtree exhibit poorer fit to the VAF data. This phenomenon was observed across all B-ALL data sets with more than 60 mutations, indicating that Pairtree's reconstructions become increasingly less reliable with more data. In contrast, Orchard's reconstructions exhibit excellent data fit irrespective of the number of mutations or samples. 

The poor data fit for Pairtree's best reconstructed tree for SJBALL022611 using all 29 samples is exemplified in Figure \ref{fig:ball-panel}c. This figure presents a heatmap of the absolute differences between the VAFs inferred by Pairtree and those implied by the bulk data for each mutation, and it shows that Pairtree's inferred VAFs significantly diverge from those implied by the bulk data, indicating a poor reconstruction. Again, in contrast, Figure \ref{fig:ball-panel}b demonstrates that Orchard's reconstructions fit the VAF data very well, as evidenced by the minimal difference between the data-implied VAFs and those inferred by Orchard.

Next, we use the phylogeny-aware clustering algorithm to infer clones from each method's best tree that was reconstructed using all of the samples for SJBALL022611. We can compare the inferred clones to the expert-defined clones using the Adjusted Rand Index (ARI). The subclones inferred from Orchard's tree exhibited an ARI of 0.96, indicating a near-perfect match with the expert-defined subclones. In contrast, the subclones inferred from Pairtree's tree were more crude, obtaining a notably lower ARI of 0.82. With the exception of two B-ALL data sets (SJBALL022609, SJBALL022610), the subclones identified by the phylogeny-aware clustering algorithm using Orchard's tree matched the expert-defined subclones as well or better than those identified using Pairtree's tree. These results are consistent with our expectations, as we believe better tree reconstructions should be more informative for guiding clone inference. We provide a more detailed discussion of the phylogeny-aware clustering results for the B-ALL data in Appendix \ref{appendix:ball-pac}.

\section*{Discussion}


Modern cancer evolution studies generate large amounts of bulk sequencing data often containing hundreds of mutations and multiple cancer samples. This vast amount of data presents a promising opportunity to improve our understanding of the evolutionary events that drive cancer progression. Until now, existing mixed sample perfect phylogeny (MSPP) reconstruction algorithms were unable to reliably reconstruct trees with more than 30 mutations. These limitations necessitated clustering mutations into subclones prior to reconstructing trees, which is an error prone process \cite{salcedo_community_2020, sun_between-region_2017, williams_identification_2016}. We introduced Orchard, a novel MSPP reconstruction algorithm that uses variant allele frequency data to build large and accurate mutation trees. We derived an approximation to the mutation tree posterior, then showed how Orchard adapts Ancestral Gumbel-Top-k sampling to efficiently sample without replacements from it. Our results demonstrate that Orchard can reliably match or beat ground-truth (or expert-derived) baselines on tree reconstruction problems with up to 300 mutations and up to 100 samples from the same cancer. Orchard can recover larger trees on simulated data, and in Appendix \ref{appendix:1000k-mutation-trees} we use Orchard to successfully reconstruct 1000-node simulated mutation trees in $\approx 26$ hours of wall clock time. Orchard's ability to reconstruct extremely large trees motivated a new strategy where the mutation tree structure is used to guide the inference of clones and clone trees. To this end, we introduced a ``phylogeny-aware" clustering algorithm that uses agglomerative clustering to infer clones and clone trees from mutation trees. 

There are a few potential areas of improvement for Orchard. First, the majority of Orchard's run time is used by the projection algorithm \cite{jia_efficient_2018} to estimate the clonal proportion matrix $U$ each time it adds a new mutation to a tree. Although the computational complexity for estimating $U$ is, at worst, quadratic in the number of mutations (i.e., rows) in $U$, the wall clock times grows quickly with increasing problem size, beam width ($k$), or branching factor ($f$). One potential improvement, because the frequency of most mutations should not be changed by introducing a new mutation into the tree, is developing a new version of the projection algorithm that only performs local updates to the $U$ matrix. Also, Orchard marginalizes over all possible ways to place a new mutation into a partially built tree, which requires an exponential number of evaluations when the tree is a star tree, i.e., a tree where all mutations are children of the root. Unless specified otherwise, Orchard will try all $2^\ell + \ell$ possible ways to place a new mutation into a star tree with $\ell$ mutations. Fortunately, under the ISA, such trees are rare and linear branching becomes more likely as the number of children for a node increases. 
So, in practice, it is rare for nodes to have large numbers of children. 

The current implementation of the ``phylogeny-aware" clustering algorithm uses a non-probabilistic merging strategy to identify clones in a mutation tree. Despite this, the algorithm already matches, or in some cases, outperforms state-of-the-art probabilistic VAF-based clustering algorithms on real data (see Appendix \ref{appendix:ball-pac}). These results suggest that the mutation tree structure is useful for guiding the inference of clones and clones trees. We anticipate that improvements to the current algorithm to make it probabilistic, so that its merging choices are made based on an appropriate likelihood function, will result in even greater performance.

We further anticipate the development of new algorithms that utilize large mutations trees reconstructed by Orchard. For example, these trees could be used to study patterns of somatic mutations, recover lineage-specific mutation signature activities, or identify patterns of metastatic spread \cite{el-kebir_inferring_2018}. The Orchard algorithm could also be adapted to reconstruct mutation trees from single-cell data, though this requires updating Orchard's noise model and sampling routine to accommodate single cell data.

\section*{Supporting information}

\paragraph*{S1 File.}
\label{appendix}
\textbf{Appendix}

\pagebreak
\makeatletter 
\renewcommand\thesection{A\@arabic\c@section}
\renewcommand\thetable{A\@arabic\c@table}
\renewcommand \thefigure{A\@arabic\c@figure}
\makeatother

\section{Calculating Orchard's Inputs}

In this section, we provide details for calculating inputs required by Orchard. The actual input file formats for Orchard are the same as Pairtree \cite{wintersinger_reconstructing_2022}, and these inputs have been described in detail in previous protocols \cite{kulman_reconstructing_2022}.

\subsection{Computing the variant read probability}\label{appendix:variant-read-probability}

The variant read probability for mutation $j$ in sample $s$, denoted as $\omega_{js} \in [0,1]$, is a correction that maps from the frequency of cells containing the mutant allele $j$ in the sample $s$ to the associated variant allele frequency of $j$. Orchard requires some basic information about the bulk sample $s$ to compute $\omega_{js}$. First, Orchard needs to know $M_{js}$, the number of copies of the mutant allele in cells in $s$ that contain it. Orchard assumes that all $j$-containing cells in $s$ have the same number of $j$ alleles. Also, we need to know $N_{js}$, the average number of copies of the genomic locus containing $j$ among the cells in $s$. Given $N_{js}$ and $M_{js}$, then
\begin{equation}\label{supeq:var-read-prob-cancerous-mut}
    \omega_{js} = \frac{M_{js}}{N_{js}}.
\end{equation}
Often cancer cells have different copy numbers at specific loci compared to normal cells. In that case, we set
\[
    N_{js} = \rho_s N^{(c)}_{js} + (1-\rho_s)N^{(h)}_{js},
\]
where
\begin{itemize}
    \item $N^{(c)}_{js}$ is the population average copy number of the locus containing the mutant allele $j$ in the cancerous cell population,

    \item  $N^{(h)}_{js}$ is the copy number of the locus containing the mutant allele $j$ in the healthy cell population. For autosomes, $N^{(h)}_{js} = 2$, and

    \item $\rho_s$ is the fraction of cells in bulk sample $s$ that are cancerous, this is also known as the purity.
    
\end{itemize}
The values $\rho_s$ and $N^{(c)}_{js}$ are outputs of CNA calling pipelines, and so should always be available. Computing $M_{js}$ can be more challenging, we suggest, due to the ISA, using $M_{js} = 1$ unless there is strong evidence that the $j$-allele has been amplified. In that case, allele-specific CNA callers provide the multiplicity of the major allele $A_{js}$ and the minor allele $B_{js}$, where $N^{(c)}_{js} = A_{js} + B_{js}$. In this case $M_{js} = A_{js}$ may be appropriate if $A_{js}$ is a whole number. However, if a locus has many CNAs across the available samples, accurately estimating $M_{js}$ can be challenging because of the possibly of subclonal changes in $j$'s multiplicity. In such scenarios, it may be necessary to exclude point mutations within genomic regions significantly affected by CNAs. For more details on estimating $M_{js}$ and $N_{js}$, please see \cite{tarabichi_practical_2021}.

\subsection{Supervariant approximation}

If Orchard is used to construct a clone tree, then it expects as input a set of mutation clusters representing individual clones. Orchard treats all mutations in the same clone as a single mutation or alternatively a ``supervariant". A ``supervariant" is a concept originally from \cite{wintersinger_reconstructing_2022} that approximates the data for a set of mutations $C$ as a single mutation. We provide a brief explanation of this approximation scheme but refer the reader to \cite{wintersinger_reconstructing_2022} for complete details.

A ``supervariant" approximation represents a group of mutations as a single mutation that can be added to a mutation tree. If $C$ is a finite set of mutations, then the ``supervariant" approximation of $C$ combines the data of each mutation $j \in C$ in each samples $s=1,\dots,m$, such that the data for $C$ in sample $s$ can be represented as $X_{Cs} = \left(a_{Cs}, b_{Cs}, \omega_{Cs}\right)$, where $a_{Cs}$, $b_{Cs}$, and $\omega_{Cs}$ are the reference read count, variant read count, and variant read probability for $C$ in sample $s$, respectively. One issue when combining the mutation data in $C$ is that it is not guaranteed that all mutations $j \in C$ have the same variant read probability in sample $s$, $\omega_{js}$. To resolve this, $\omega_{Cs} = \frac{1}{2}$ is fixed for all samples $s = 1,\dots,m$, and the data for all mutations $j \in C$ is scaled as follows:

\begin{align*}
    T_{js} &= a_{js} + b_{js} \\
    \Tilde{T}_{js} &= 2\omega_{js}T_{js} \\ 
    \Tilde{b}_{js} &= min(b_{js}, \Tilde{T}_{js}) \\ 
    \Tilde{a}_{js} &= \Tilde{T}_{js} - \Tilde{b}_{js} \\ 
    \Tilde{\omega}_{js} &= \frac{1}{2}
\end{align*}
After scaling the data for each mutation $j \in C$ for all samples $s = 1,\dots,m$, we can compute the values for $X_{Cs}$ as follows:

\begin{align*}
    X_{Cs} &= \left\{a_{Cs}, b_{Cs}, \omega_{Cs}\right\} \\ 
    a_{Cs} &= round\left(\sum_j{\Tilde{a}_{js}}\right) \\ 
    b_{Cs} &= round\left(\sum_j{\Tilde{b}_{js}}\right) \\ 
    \omega_{Cs} &= \frac{1}{2}
\end{align*}
There are a few problems with the ``supervariant" approximation. First, using the ``supervariant" approximation results in a slightly different reconstruction problem than a mutation tree reconstruction problem, where the perfect phylogeny matrix $B$ and the clonal proportion matrix $U$ are inferred for the ``supervariants" instead of the mutations. Also, grouping mutations with different variant read probabilities means that we assume mutations co-occur even though they are in genomic regions that vary in copy number. There is, in fact, uncertainty about whether or not these mutations are part of the same clone. Since Orchard can reconstruct extremely large trees, we could instead use a \textit{relaxed ``supervariant" approximation}, where pre-defined mutation clusters are split apart such that only mutations with the same variant read probability appear in the same clone. This relaxation would remove some of the pitfalls introduced by the ``supervariant" approximation. For simplicity, we chose to use the ``supervariant" approximation as described here.

\section{Factorized approximation to the mutation tree posterior}\label{appendix:computing-factorized-approximate-posterior}

In this section, we provide further details on the factorized approximation to the mutation tree posterior presented in Eq \ref{eq:approximate-factorized-posterior}.

\subsection{Computing a mutation order}\label{appendix:mutation-ordering}

The factorized approximate posterior, $Q^{\pi}(B|D)$, relies on a mutation order $\pi$. We can design an order of mutations such that no mutation comes before its ancestors by ensuring the ordering adheres to the \textit{sum constraint} \cite{jiao_inferring_2014, el-kebir_reconstruction_2015} implied by the ISA. For a mutation order to adhere to the sum constraint it must be the case that:

\begin{equation*}
     u \succ v : \widehat{F}_{us} \geq \widehat{F}_{vs} \hspace{0.1cm} \forall s,
\end{equation*}
i.e., mutation $u$ must come before mutation $v$ in the order if its frequency is at least as large in each sample. If the frequencies of $u$ and $v$ are the same across all samples, then either ordering $u \succ v$ or $v \succ u$ adheres to this constraint. If a mutation ordering can be found that agrees with the sum constraint, then each mutation could be added to the tree as a leaf node. However, $\widehat{F}$ often contains noise, and it's possible that no such order can be found. Instead, we relax the sum constraint and find a mutation order that adheres to the $\widehat{F}$-sum constraint:

\begin{equation*}
    u \succ v : \gamma_u \geq \gamma_v,
\end{equation*}
where $\gamma_w = \sum_{s=1}^m\widehat{F}_{ws}$. This order can always be found, regardless of whether or not $\widehat{F}$ contains noise, and if $\widehat{F}$ is noise free and the data adheres to the ISA, then this order will be the same as that which adheres to the sum constraint. Orchard sets $\pi$ according to the $\widehat{F}$-sum constraint, and since this ordering does not guarantee that mutations come before their ancestors, Orchard marginalizes over all possible placements for a mutation, both as internal nodes and leaf nodes. 

\begin{figure}[ht]
\begin{center}
\includegraphics[scale=0.28]{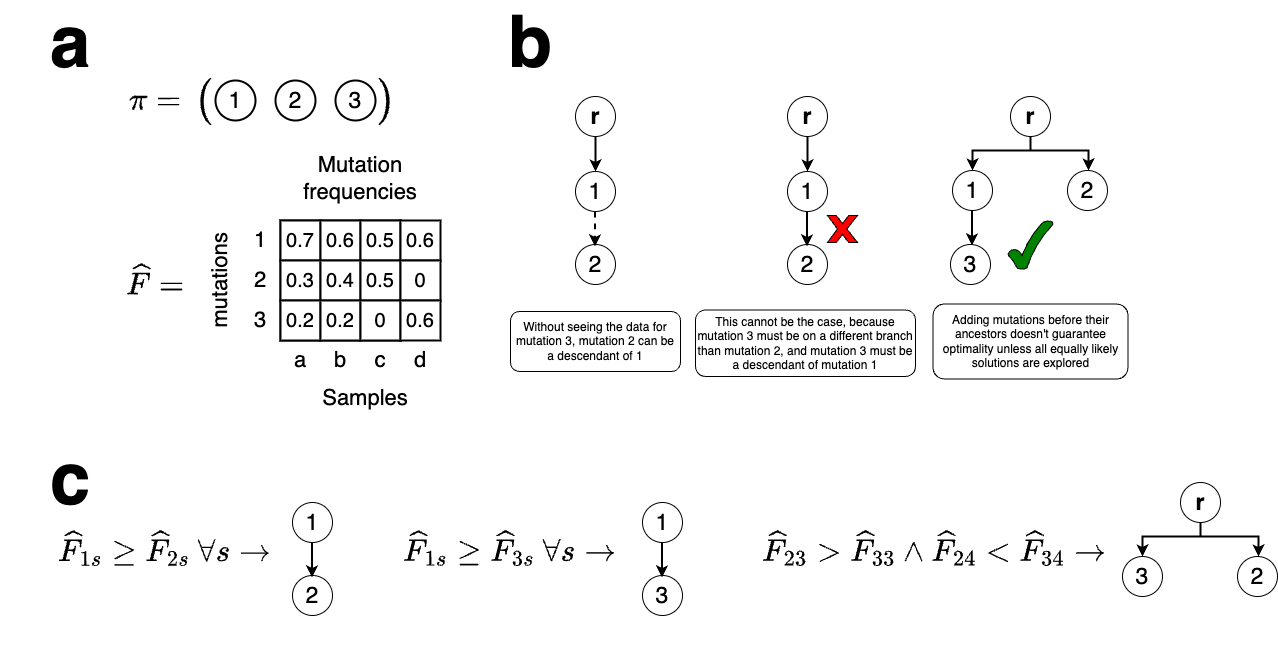}
\end{center}
\caption{Counterexample where $Q^{\pi}(B|D)$ may fail to find an optimal tree even if mutations are added before their ancestors. \textbf{a. } The mutation order $\pi$ guarantees that mutations are added before their ancestors. \textbf{b. } Adding mutation $2$ as a descendant of $1$ adheres perfectly to the ISA, but if the data for mutation $3$ is observed then it must be the case that mutation $1$ and $2$ are on separate branches. \textbf{c. } The breakdown of the plausible pairwise relationships between each pair of mutations $\left\{(1,2), (1,3), (2,3)\right\}$. There's only one possible mutation tree structure implied by these pairwise relationships, and $Q^{\pi}(B|D)$ may fail to recover it.}
\label{fig:mutation-order-counterexample}
\end{figure}

In many cases, the $\widehat{F}-$sum mutation order guarantees that no mutation is placed in the tree before its ancestors. If this is the case, then $Q^{\pi}(B|D)$ could accurately approximate $P(B|D)$, since observing $D^{(\ell+1)}$ alone is adequate for constructing an optimal partial tree. However, there are counter examples where, even if mutations are added to the tree before their ancestors, the partial tree structure precludes the discovery of the optimal tree. Figure \ref{fig:mutation-order-counterexample} shows an example where it's necessary to observe the data for the last $n - \ell + 1$ mutations in order to build an optimal partial tree.

\subsection{Alternative techniques for sampling from the approximate posterior}\label{appendix:alternative-sampling-methods}

In this section, we discuss alternative techniques to sample from $Q^{\pi}(B|D)$. These techniques include a greedy approach and categorical sampling.

\subsubsection{Greedy sampling}\label{sec:greedy-sampling}
Let $t^{(\ell)}$ be a partial tree with a genotype matrix $B^{(\ell)}$ and log probability $\phi = \text{log}Q^{\pi}(B^{(\ell)}|D^{(\ell)})$. The \textit{greedy} extension of $t^{(\ell)}$ is obtained with

\begin{equation}\label{eq:greedy-sequences}
    t^{\wedge(\ell+1)}= \argmax_{i \in \{1,2,\dots,f\}}\phi_i.
\end{equation}
This can be applied repeatedly down the search tree to obtain a greedy sample from $Q^{\pi}(B|D)$. For many MSPP problems, there will be multiple equally likely solutions, and it is desirable to obtain some or all of these solutions. It's possible to improve this estimate by modifying Eq \ref{eq:greedy-sequences} to obtain the top-$k$ greedy extensions:

\begin{equation}\label{eq:top-k-greedy-sequences}
    t_1^{\wedge(\ell+1)}, t_2^{\wedge(\ell+1)}, \dots, t_k^{\wedge(\ell+1)}= \argtopk \phi_i,
\end{equation}
where $\argtopk$ selects the top-$k$ extensions of $t^{(\ell)}$ according to their log probabilities. Repeatedly applying Eq \ref{eq:top-k-greedy-sequences} down the search tree will obtain the top-$k$ greedy samples. Although this approach generates multiple unique samples, it is deterministic and there are no guarantees about the optimality of these samples.

\subsubsection{Categorical Sampling}

Some of the pitfalls of a greedy approach can be mitigated by \textit{sampling} which extensions to keep. First, the set of extensions of $B^{(\ell)}$ must be treated as a categorical distribution, where each $B^{(\ell+1,i)}$ corresponds to a unique choice, and the probability it is chosen, i.e., $P(I = i)$, is proportional to its likelihood:

\begin{equation}\label{eq:categorical-sampling}
I \sim \text{Categorical}\left(\frac{\text{exp}(\phi_{i})}{\displaystyle\sum_{j = 1}^f\text{exp}(\phi_{j})}, i \in \{1,2,\dots,f\}\right).
 \end{equation}
Sampling helps prevent the search from becoming trapped in local maxima. Running the search multiple times may yield multiple unique mutation trees, but each run samples with replacement, so this is not guaranteed. An alternative is to draw multiple samples using Eq \ref{eq:categorical-sampling} and \textit{reject} duplicates until $k$ unique extensions are obtained. Repeatedly applying this technique will yield of sample of $k$ complete mutation trees without replacement from $Q^{\pi}(B|D)$. However, this approach can become problematic and expensive if there are numerous extensions or if the categorical distribution over these extensions has low entropy. In such cases, a very large number of duplicate samples might be drawn before $k$ distinct ones are obtained \cite{kool_stochastic_2019, kool_ancestral_2020}.

\section{Extending a partial mutation tree}\label{appendix:sequence-extension}

Orchard extends a partial mutation tree $t^{(\ell)}$ by adding a new mutation. There are two equivalent ways to think about the addition of a new mutation. First, we can add a new row $\va^{(\ell+1)}$ and column $\vd^{(\ell+1)}$ to the corresponding genotype matrix representation of $t^{(\ell)}$, $B^{(\ell)}$, then enumerate all possible ways to populate $\va^{(\ell+1)}$ and $\vd^{(\ell+1)}$ such that $B^{(\ell+1)} = \left[\begin{matrix}B^{(\ell)}, \va^{(\ell+1)};\vd^{(\ell+1)},1\end{matrix}\right]$ remains a perfect-phylogeny-compatible genotype matrix. Alternatively, we can add a mutation $v$ to $t^{(\ell)}$ by choosing parent node $u \in V^{(\ell)}$ for $v$, and then deciding which of $u$'s children in $t^{(\ell)}$ become the children of $v$. All placements of $v$ in $t^{(\ell)}$ can be considered by enumerating through all combinations of parents for $v$ and which mutations become its children. As such, the number of valid ways to add $v$ to $t^{(\ell)}$ is $\sum_{u \in V^{(\ell)}} 2^{|ch(u)|}$, where $ch(v) \subset V^{(\ell)}$ is the set of children of $u$ in $t^{(\ell)}$. In the worst-case scenario, the mutation tree resembles a star tree where all mutations are children of the root. In such cases, the number of possible ways to add $v$ to $t^{(\ell)}$ amounts to $2^{\ell } + \ell$. For example, if $u = r$ is the root which has $\ell$ children, then there are $\sum_{\ell^{\prime}=1}^{\ell}{\binom{\ell}{\ell^{\prime}}} = 2^{\ell}$ total number of placements where $v$ is a child of $u$, and then it is a binary decision of whether or not $v$ is the parent of each of the $\ell$ children of $u$. Additionally, there are $\ell$ other possible placements of $v$: it can be a child of each of the $\ell$ mutations in $t^{(\ell)}$, but $v$ is not a parent of any mutations. While it's conceivable that the VAF data suggests such a tree structure, it's highly improbable if the point mutations exhibit non-zero variant allele frequencies in one or more samples. As the number of children of $u$ with non-zero VAFs increases, the likelihood of violating the \textit{sum constraint} (Definition \ref{supdef:F-prop-2}) increases.

\subsection{Heuristic for computing the likelihood of an extension}\label{appendix:placing-mutations}

There can be numerous extensions of $t^{(\ell)}$, many of which will have near zero probability under the approximate posterior $Q^{\pi}(D|B)$. Evaluating $Q^{\pi}(D|B)$ for each of these extensions requires computing the clonal proportion matrix, which can be very time consuming. Instead, we can quickly approximate which extensions should be kept using a special function, $H(t^{(\ell)},f)$. This function selects $f$ extensions of $t^{(\ell)}$ to keep, for which we can then approximate the clonal proportion matrix and properly evaluate them under $Q^{\pi}(D|B)$. 

We'll now describe how $H(t^{(\ell)},f)$ selects these extensions. First, we show that each placement of $v$ into $t^{(\ell)}$ can be evaluated by how well each placements adheres to the ISA. Next, we show how to compute a probability distribution over these placements.

\subsubsection{Defining constraints on \texorpdfstring{$F_{vs}$}{Lg} given the placement of \texorpdfstring{$v$}{Lg}}

Under the ISA, the mutation frequency matrix $F$ has three important properties in relation to each mutation $u \in V^{(\ell)}$ for all samples $s = 1,\dots,m$:

\begin{definition}[$F$ Properties]\label{supdef:F-properties}
\mbox{}\vspace{0.1cm}
\begin{enumerate}
    \item $0 \leq F_{us} \leq 1$ \label{supdef:F-prop-1}
    \item $F_{us} \geq \sum_{w \in ch(u)}{F_{ws}}$ (sum constraint)  \label{supdef:F-prop-2}
    \item $F_{us} = U_{us} + \sum_{w \in D(u)}{U_{ws}}$  \label{supdef:F-prop-3}
\end{enumerate}
\end{definition}
where $D(u)$ denotes the set of all descendants of $v$ according to the tree structure defined by $t^{(\ell)}$. The value $U_{us}$ denotes the clonal proportion of mutation $u$ in sample $s$ which is defined as 

\begin{equation}\label{supeq:eta-definition}
U_{us} = F_{us} - \sum_{w \in ch(u)}{F_{ws}},
\end{equation}
where $ch(u)$ is the set of mutations that are children of $u$ in $t^{(\ell)}$. Another important relationship we assume for $F_{us}$ is that 

\begin{equation}\label{eq:vaf-to-frequency}
    F_{us} = \frac{\lambda_{us}}{\omega_{us}},
\end{equation}
where $\lambda_{us}$ is the variant allele frequency of mutation $u$ in sample $s$, and $\omega_{us}$ is the variant read probability of mutation $u$ in sample $s$. This is equivalent to Eq \ref{eq:vaf-translation}.

All extensions of $t^{(\ell)}$ correspond to unique placements of a new mutation $v \in V \setminus V^{(\ell)}$ into the tree. These extensions can be enumerated by selecting a parent $u \in V^{(\ell)}$, and choosing which of $u$'s children become the children of $v$. Once $v$ is placed in $t^{(\ell)}$, it must also be the case that $F_{vs}$ adheres to the constraints in Definition \ref{supdef:F-properties}. We can use this fact to compute a probability distribution over all extensions of $t^{(\ell)}$ based on how well the VAF data for $v$ supports the constrained values of $F_{vs}$.

Let $u$ be the parent of $v$ and $\chi \subseteq ch(u)$ be a set of mutations that are now the children of $v$. We can use Definition \ref{supdef:F-properties}, Property \ref{supdef:F-prop-2}, to define a set of constraints on $F_{vs}$. Since $v$ is now the parent of the mutations in $\chi$, it must be the case that:

\begin{equation*}
F_{vs} \geq \sum_{i \in \chi}{F_{is}}.
\end{equation*}
At the same time, since $v$ is a child of $u$, it must also be the case that 

\begin{equation*}
F_{us} \geq F_{vs} + \sum_{j \in ch^*(u)}{F_{js}},
\end{equation*}
where $ch^*(u) = ch(u) \setminus \chi$ is the set of mutations that remain as children of $u$. We can combine these two inequalities to obtain the following constraints on $F_{vs}$:

\begin{equation}\label{supeq:u-v-initial-constraints}
F_{us} - \sum_{j \in ch^*(u)}{F_{js}} \geq F_{vs} \geq \sum_{i \in \chi}{F_{is}}.
\end{equation}
We can then use Equation \ref{eq:vaf-to-frequency} to rewrite eq \ref{supeq:u-v-initial-constraints} as

\begin{align}\label{supeq:u-v-q-constraints}
F_{us} - \sum_{j \in ch^*(u)}{F_{js}} &\geq \frac{\lambda_{vs}}{\omega_{vs}} \geq \sum_{i \in \chi}{F_{is}}, \nonumber \\
\omega_{vs}\left(F_{us} - \sum_{j \in ch^*(u)}{F_{js}}\right) &\geq \lambda_{vs} \geq \omega_{vs}\left(\sum_{i \in \chi}{F_{is}}\right),
\end{align}
where $\lambda_{vs}$ is the VAF of mutation $v$ in sample $s$ and $\omega_{vs}$ is the variant read probability of mutation $v$ in sample $s$. It's clear from Eq \ref{supeq:u-v-q-constraints} that if $\chi \neq \emptyset$, then under the ISA $\lambda_{vs}$ needs to fall below $\omega_{vs}\left(F_{us} - \sum_{j \in ch^*(u)}{F_{js}}\right)$, and above $\omega_{vs}\left(\sum_{i \in \chi}{F_{is}}\right)$.

 Let's consider the case where $v$ is a child of $u$ and $\chi = \emptyset$. We can use Definition \ref{supdef:F-properties}, Property \ref{supdef:F-prop-2}, to derive the constraint:

 \begin{equation*}
 F_{us} \geq \sum_{i \in ch(u)}{F_{is}} + F_{vs}.
 \end{equation*}
 By rearranging terms we obtain the following:

\begin{align}  
F_{us} - \sum_{i \in ch(u)}{F_{is}} &\geq F_{vs}\nonumber \\ 
U_{us} &\geq F_{vs}\label{supeq:eta-geq-F} \\ 
U_{us} &\geq \frac{\lambda_{vs}}{\omega_{vs}}\nonumber \\ 
\omega_{vs}U_{us} &\geq \lambda_{vs}\label{supeq:u-v-noq-constraints}
\end{align}
The constraint on the left hand side in Eq \ref{supeq:eta-geq-F} is obtained by the definition of the clonal proportion $U_{us}$ (Eq \ref{supeq:eta-definition}). We again use Eq \ref{eq:vaf-to-frequency} to obtain our final result in Eq \ref{supeq:u-v-noq-constraints}.

Finally, we can use the constraints in Eq \ref{supeq:u-v-q-constraints} and Eq \ref{supeq:u-v-noq-constraints} to define probabilities for the parent/child relationships for $v$:

\begin{align}
p\left(u, v, \chi| \chi \neq \emptyset\right) &= Pr\left[\omega_{vs}\left(F_{us} - \sum_{j \in ch^*(u)}{F_{js}}\right) \geq \lambda_{vs} \geq \omega_{vs}\left(\sum_{i \in \chi}{F_{is}}\right)\right]\label{supeq:u-v-q-prob} \\ 
p\left(u,v,\chi|\chi=\emptyset\right) &= Pr\left[\omega_{vs}U_{vs} \geq \lambda_{vs}\right]\label{supeq:u-v-noq-prob}
\end{align}
It's important to note that in practice Eq \ref{supeq:u-v-q-prob} is computed for all $\chi \in \mathcal{P}(ch(u))$, where $\mathcal{P}(ch(u))$ is the \textit{power set}, or set of all subsets, of $ch(u)$. Note that the size of $\mathcal{P}(ch(u))$ grows exponentially with the number of children of $u$. Orchard has a parameter that can be set by the user which limits the number of placements scored using Eq \ref{supeq:u-v-q-prob}. This parameter is called the \textit{max placements} parameter, and it provides an upper limit on the number of $u$'s children that $v$ can parent. For example, if the \textit{max placements} is set to 5, then $H(t^{(\ell)}, f)$ will consider all possible placements where $v$ is a child of $u$, and $v$ is the parent of any combination of $i$ children of $u$, where $i = 0,\dots,5$. As a result, $H(t^{(\ell)},f)$ will consider $\sum_{i=0}^5{\binom{|ch(u)|}{i}}$ possible placements of $v$ when it's a child of $u$.

\subsubsection{Computing the posterior over \texorpdfstring{$\lambda_{vs}$}{Lg}}

In this section, we show how to compute the probabilities in Eq \ref{supeq:u-v-q-prob} and Eq \ref{supeq:u-v-noq-prob}. 

Under a binomial sampling model, we assume that our observed variant allele counts $b$ are binomially distributed according to some success probability $\lambda$, and that we've mapped a total of $N = a + b$ reads to that locus. We place no prior beliefs on the distribution of $\lambda$, and therefore use a uniform $Beta$ prior on its distribution. Under these assumptions, the posterior distribution of $\lambda$ for some mutation $v$ in sample $s$ is defined as follow:

\begin{align*}
b | \lambda_{vs} &\sim Binom(N_{vs}, \lambda_{vs}) \\
\lambda_{vs} &\sim Beta(\alpha_0, \beta_0)  \\
\lambda_{vs} | (b = b_{vs}) &\sim Beta(\alpha_0 + b_{vs}, \beta_0 + N_{vs} - b_{vs}),
\end{align*}
where $N_{vs} = a_{vs} + b_{vs}$, and the \textit{Beta prior} is uniform, so $\alpha_0 = \beta_0 = 1$. Because the $Beta$ is conjugate to the the binomial, the posterior distribution of $\lambda_{vs}$ is also a $Beta$ distribution, in this case with $\alpha=1+b_{vs}$ and $\beta=1+a_{vs}$. We can rewrite the posterior in terms of solely $\lambda_{vs}$ and $X_{vs}$ as follows:

\begin{align}\label{supeq:posterior-lambda}
    p(\lambda_{vs}|X_{vs}) &\propto p(X_{vs}|\lambda_{vs})p(\lambda_{vs}) \nonumber \\
    &\propto f(\lambda_{vs}, \alpha, \beta)
\end{align}
where $X_{vs} = (a_{vs}, b_{vs}, \omega_{vs})$ and $f(\lambda, \alpha, \beta)$ is the probability density function of the Beta Distribution. If we denote $y_s = \omega_{vs}\left(\sum_{i \in \chi}{F_{is}}\right)$ and $z_s = \omega_{vs}\left(F_{us} - \sum_{j \in ch^*(u)}{F_{js}}\right)$, then we can define Eq \ref{supeq:u-v-q-prob} as follows:

\begin{align}
p\left(u, v, \chi| \chi \neq \emptyset\right) &= Pr\left[z_s \geq \lambda_{vs} \geq  y_s\right] \nonumber\\
&= \int_{y_s}^{z_s}{f(\lambda_{vs}, \alpha, \beta)d\lambda_{vs}} \nonumber\\
&= \int_{y_s}^{z_s}{\frac{1}{B(\alpha, \beta)}\lambda_{vs}^{\alpha}(1-\lambda_{vs})^{\beta}d\lambda_{vs}} \nonumber\\ 
&= \frac{1}{B(\alpha, \beta)}\int_{y_s}^{z_s}{\lambda_{vs}^{\alpha}(1-\lambda_{vs})^{\beta}d\lambda_{vs}} \nonumber\\ 
  &=  I_{z_s}(\alpha,\beta) - I_{y_s}(\alpha,\beta)\label{supeq:u-v-q-betainc}
\end{align}
where $I_{x}(\alpha,\beta)$ is the \textit{regularized incomplete beta function}. If we let $z'_s = \omega_{vs}U_{vs}$, then we can perform a similar process to define Eq \ref{supeq:u-v-noq-prob} as

\begin{align}
p\left(u,v,\chi|\chi=\emptyset\right) &= Pr\left[z'_s \geq \lambda_{vs}\right] \nonumber\\
&= \int_{0}^{z'_s}{f(\lambda_{vs}, \alpha, \beta)d\lambda_{vs}} \nonumber\\
&= \frac{1}{B(\alpha, \beta)}\int_{0}^{z'_s}{\lambda_{vs}^{\alpha}(1-\lambda_{vs})^{\beta}d\lambda_{vs}} \nonumber\\ 
  &=  I_{z'_s}(\alpha,\beta)\label{supeq:u-v-noq-betainc}
\end{align}
It's possible that $y_s$ and $z_s$ can be very close in Eq \ref{supeq:u-v-q-betainc}. This might occur when mutations are part of the same clone, and therefore have very close mutation frequencies across all samples. In this case, we would like Orchard to place these mutations together in the tree, however, the closeness of the mutation frequencies may make it so the ideal placements have near zero probability. To mitigate this problem, we widen the range of integration between $y_s$ and $z_s$ by adjusting them using binomial proportion confidence intervals. In particular, we use the Wald interval to adjust $y_s$ and $z_s$:

\begin{align*}
    \hat{z}_s = z_s + \frac{2.576}{\sqrt{N_{vs}}}\sqrt{\widehat{\lambda}_{vs}(1-\widehat{\lambda}_{vs})}, \\ 
    \hat{y}_s = y_s - \frac{2.576}{\sqrt{N_{vs}}}\sqrt{\widehat{\lambda}_{vs}(1-\widehat{\lambda}_{vs})}. \\ 
\end{align*}
The value $\widehat{\lambda}_{vs}$ is the variant allele frequency of mutation $v$ in sample $s$ that's implied by the data, which can be calculated using the formula presented in Section \ref{sec:orchard-inputs}. The value ``$2.576$" is the target error rate that corresponds to the 99th percentile. If $y_s = z_s = \widehat{\lambda}_{vs}$, then $\hat{z}_s$ and $\hat{y}_z$ would be the 99th percentile upper confidence limit and 99th percentile lower confidence limit for $\widehat{\lambda}_{vs}$, respectively. Under these conditions the integral in Eq \ref{supeq:u-v-q-betainc} will evaluate to nearly 1, meaning this placement agrees perfectly with the data for mutation $v$ in sample $s$. 

\subsubsection{Choosing a branching factor}

The function $H(t^{(\ell)},f)$ quickly evaluates all possible extensions of $t^{(\ell)}$ and returns the indices of the top-$f$ extensions under its approximation. We can assign a ``rank" to each extension using the sorted order of the likelihoods computed by $H(t^{(\ell)},f)$: the extension with the largest likelihood is assigned rank 1, the second largest as rank 2, etc.. As Orchard builds trees, we can record these ranks and evaluate what we should set $f$ to such that we evaluate $Q^{\pi}(B|D)$ as few times as possible, while not impacting the accuracy of Orchard's reconstructions. 

We empirically determined the value of the branching factor $f$ to use in our experiments using a validation set. We ran Orchard on this validation set with parameters that allow it to evaluate  $Q^{\pi}(B|D)$ for all possible extensions by setting $f = \infty$. This does not result in a complete exhaustive search of all possible trees (which would be computationally intractable for the tree sizes in the validation set) because we limit the number of partial trees to keep track of by setting the beam width parameter $k$. We chose to run Orchard with two different beam widths $k = 1$ and $k = 10$, along with $f = \infty$. We decided to use 200 simulated bulk DNA cancer phylogenies originally from \cite{myers_calder_2019} as a validation set. These simulations mimic whole exome sequencing, with each dataset containing 100-200 mutations, 5 samples, and a read depth of 200x.

\begin{figure}[ht]
\begin{center}
\includegraphics[width=350px,height=250px]{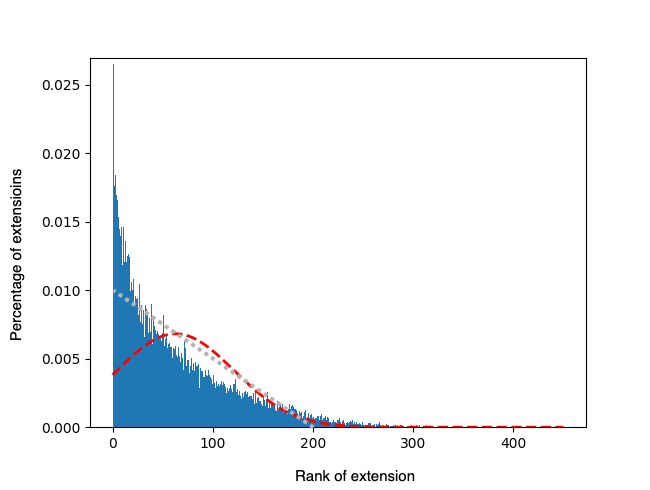}
\end{center}
\caption{The percentage of extensions of each rank that led to the tree with the largest likelihood across all data sets in the validation set. The validation set consisted of 200 simulated cancers originally from \cite{myers_calder_2019}. The grey dotted line represents the percentage of extensions that we would expect for each rank if they were randomly chosen, assuming only a linear tree structure. The red dotted line is a Gaussian distribution fit to the rank data.}
\label{supfig:rank_k1}
\end{figure}

Figure \ref{supfig:rank_k1} shows the percentage of ranks that led to the tree with the largest likelihood across all datasets in the validation set. Setting $k=1$ and $k=10$ results in very similar distributions of ranks; thus, we only show the results for $k=1$. The 30th percentile rank is 21, the 50th percentile is 46, and the 80th percentile is 109. To find the optimal extension 80\% of time, we need to evaluate $Q^{\pi}(B|D)$ for the top 100 extensions. While this is computationally tractable, in practice we found on the validation set that the results were still very good and 5x faster when considering only the top 20 extensions. Based on these findings, we chose to limit $f=20$ in our experiments. 

The results in Figure \ref{supfig:rank_k1} support that $H(t^{(\ell)},f)$ is a relatively good approximation. However, further improvements to the heuristic could dramatically decrease run times and improve reconstructions. Additionally, it is possible to adapt this heuristic into its own MSPP reconstruction method, with the complete trees output by the method being formally evaluated using $P(B|D)$.

\subsection{Derivations}\label{appendix:approximate-posterior-derivations}


We now show that the relation in Eq \ref{eq:factorized-posterior} holds true. Let $B \in \left\{0,1\right\}_{n \times n}$ be a binary genotype matrix, where each entry $B_{ij} \in \left\{0,1\right\}$ is a binary random. We denote the set of all of these binary random variables with $A_B = \left\{B_{ij} | B_{ij} \text{ is an element of } B\right\}$. The joint posterior over this set is equivalent to Eq \ref{eq:posterior}:

\begin{equation*}
    P(B|D) = P(A_B|D)
\end{equation*}
If $A_{B_{:j}} = \left\{B_{1j}, B_{2j}, \dots, B_{nj}\right\}$ is the set of binary random variables associated with the $j$-th column of $B$, then the joint posterior can be rewritten as:

\begin{equation*}
    P(A_B|D) = P(A_{B_{:1}}, A_{B_{:2}}, \dots, A_{B_{:n}}|D) = \prod_{j=1}^{n-1}P(A_{B_{:j+1}}| A_{B_{:1}},\dots,A_{B_{:j}}, D),
\end{equation*}
which is a straight forward application of the chain rule of probability. We can perform a similar factorization using arbitrary subsets of $A_B$. Consider  $B^{(\ell+1)} = \left[\begin{matrix}B^{(\ell)}, \va^{(\ell+1)};\vd^{(\ell+1)},B_{\ell+1,\ell+1} \end{matrix}\right]$, where:

\begin{itemize}
    \item $\va^{(\ell+1)}$ represents the first $\ell$ elements of the $(\ell+1)$-th column of $B$
    \item $\vd^{(\ell+1)}$ represents the first $\ell$ elements of the $(\ell+1)$-th row of $B$
    \item $B^{(\ell)}$ is a submatrix containing the first $\ell$ rows and columns of $B$
    \item $B_{\ell+1,\ell+1}$ is the $(\ell+1)$-th entry on the diagonal of $B$.
\end{itemize} 
Let $A_{\va^{(\ell+1)}}, A_{\vd^{(\ell+1)}}, A_{B^{(\ell)}}, A_{B_{\ell+1,\ell+1}}$ denote the sets of binary random variables corresponds to these parts of $B$, then we can rewrite the factorization again as:

\begin{equation*}
    P(B|D) = P(A_B|D) = \prod_{\ell=1}^{n-1}P(A_{\va^{(\ell+1)}}, A_{\vd^{(\ell+1)}}, A_{B_{\ell+1,\ell+1}}|  A_{B^{(\ell)}},D) = \prod_{\ell=1}^{n-1}P(B^{(\ell+1)}| B^{(\ell)}, D),
\end{equation*}
which shows that Eq \ref{eq:factorized-posterior} holds true.

\section{Phylogeny-aware clustering}\label{appendix:phylogeny-aware-clustering}

The ``phylogeny-aware" clustering algorithm we propose consists of two steps: $(1)$ agglomerative clustering, and $(2)$ model selection. We will now describe these steps in detail.

\subsection{Agglomerative clustering}

The phylogeny-aware clustering algorithm we propose is an \textit{agglomerative clustering algorithm}. Agglomerative clustering, also known as \textit{hierarchical agglomerative clustering}, is a greedy clustering method that iteratively merges pairs of nodes in a graph until the graph consists of only a single node \cite{mullner_modern_2011}. Generally, agglomerative clustering methods have a worst case time complexity of $O(n^3)$ \cite{mullner_modern_2011}. In order to determine which pair of nodes should be combined at each step during agglomerative clustering, a \textit{cluster linkage} criterion is used to evaluate the dissimilarity between all adjacent nodes in the graph. Since adjacent nodes $u$ and $v$ contain one or more objects (mutations), the cluster linkage criterion is generally a function of the dissimilarity between each pair of objects (mutations) $(i,j)$ where $i \in u, j \in v$. The phylogeny-aware clustering algorithm iteratively joins nodes based on minimizing the linkage criterion:

\begin{equation*}
\text{min}_{u,v}{d(u,v)},
\end{equation*}
where $d(u,v)$ is a distance function. In our case, we chose to use Ward's method for the linkage criterion because it is a standard for agglomerative clustering algorithms, where:

\begin{equation*}
d(u,v) = \frac{n_un_v}{n_u + n_v}\lVert \bar{F}_u - \bar{F}_v \rVert^2,
\end{equation*}
where $n_u$ is the number of mutations associated with node $u$, $\bar{F}_u$ is the average mutation frequency of the mutations contained in node $u$, and $\lVert \cdot \rVert$ is the Euclidean norm. 

The phylogeny-aware clustering algorithm performs agglomerative clustering on a directed rooted tree, and so special care must be taken when merging nodes. Importantly, we at first only consider joining adjacent nodes that are on the same \textit{linear segment} until all such pairs are exhausted. We define a linear segment as a linear sequence of nodes, where each node has exactly 1 or 0 child nodes. Nodes that are not on the same linear segment are distinguished by a \textit{speciation} or \textit{branching} event, representing the occurrence of distinct clones. Mutations that occur after a branching event are unique to that clone and all its subclones. However, we relax this constraint and allow the algorithm to consider merges that either join distinct branches or join a child node to a parent that has more than one child once there are no more linear segments. Thus, the algorithm tries to adhere to the phylogenetic constraints when clustering until no merges can be performed without violating them. This approach allows the user to consider clone trees containing anywhere from $1$ to $n$ clones. 

\begin{figure}[ht]
\begin{center}
\includegraphics[scale=0.3]{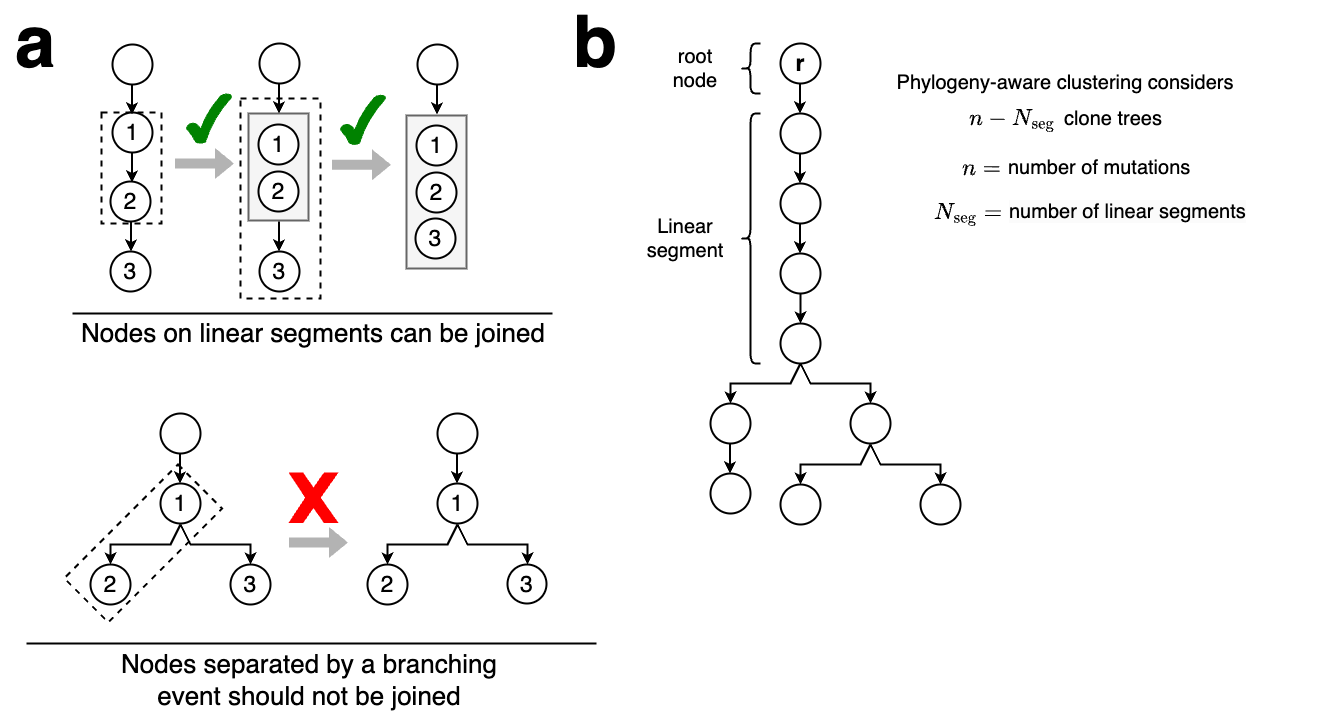}
\end{center}
\caption{Overview of how agglomerative clustering ideally operates on a tree. \textbf{a. } Phylogenetic constraints imply that only adjacent nodes on the same linear segment should be considered for merging. Merging nodes that are separated by a branching event violates the phylogenetic constraints. Dotted rectangles enclose the nodes considered for joining. \textbf{b. } The number of clone trees that can be resolved without violating phylogenetic constraints can be calculated using the formula: $n - N_{\text{seg}}$, where $n$ is the number of mutations in the mutation tree, and $N_{\text{seg}}$ is the number of linear segments in the tree. For the tree shown in \textbf{b}, $n = 9$, and $N_{\text{seg}} = 5$, therefore, there are 4 clones trees that can be resolved with agglomerative clustering if phylogenetic constraints aren't violated.}
\label{supfig:phylogeny-aware-clustering-ops}
\end{figure}

At each iteration, the phylogeny-aware clustering algorithm merges two adjacent nodes. Consequently, at iteration $i$, the tree consists of $n - i$ clones. If we restrict the phylogeny-aware clustering algorithm to only merge nodes on linear segments, then the minimum number of clones resolvable in a mutation tree is defined by the number of linear segments, $N_{\text{seg}}$. Therefore, if $N_{\text{seg}}$ is the number of linear segments in a mutation tree and $n$ is the number of mutations, then there are at most $n - N_{\text{seg}}$ clones trees that can be resolved with agglomerative clustering without violating phylogenetic constraints. The phylogenetic constraints that should be adhered to when performing agglomerative clustering on a tree are illustrated in Figure \ref{supfig:phylogeny-aware-clustering-ops}a. The concept of a linear segment, calculating the minimum number of clones in a tree, and calculating the maximum number of clone trees resolvable without violating phylogenetic constraints are illustrated in Figure \ref{supfig:phylogeny-aware-clustering-ops}b.

\subsection{Estimating the cluster variant allele frequency}

The likelihood of each clustering with $c = 1,\dots,n$ clones is calculated using Eq \ref{eq:pac-likelihood}. In order to evaluate the likelihood, the variant allele frequency for each clone $i$ in sample $s$, $\lambda_{is}$, needs to be estimated. The mutation frequencies, $F$, and the VAF data, $D$, are used estimate $\lambda_{is}$ as follows:

\begin{equation}\label{eq:pac-vaf-estimate}
    \lambda_{is} = \frac{\sum_{j|Z_{ji}=1}F_{js}\omega_{js}N_{js}}{\sum_{j|Z_{ji}=1}N_{js}}.
\end{equation}
The quantity $F_{js}\omega_{js}N_{js}$ estimates the number of variants reads expected to be observed in sample $s$ given that $\lambda_{js} = F_{js}\omega_{js}$ and a total of $N_{js}$ reads were mapped to the locus containing mutation $j$. The numerator in Eq \ref{eq:pac-vaf-estimate} is the sum of the expected number of variant reads in sample $s$ for each mutation assigned to clone $i$, while the denominator is the total number of reads mapped to each of the loci containing a mutation assigned to clone $i$. The relationship $\lambda_{js} = F_{js}\omega_{js}$ follows directly from Eq \ref{eq:vaf-to-frequency}.

\subsection{Model selection for phylogeny-aware clustering}

The phylogeny-aware clustering method outputs $n$ different models as a result of its agglomerative clustering scheme. Each model contains a unique number of clusters from $1$ to $n$. We can then either manually select the number of clusters believed to be in the data set, or we can use a model selection criterion to choose the number of clusters. One of the most well known model selection criteria is the Bayesian Information Criterion (BIC) \cite{neath_bayesian_2012}. The BIC chooses a model from a finite set of proposed models based on a penalized likelihood of the data given the model and its parameters. The BIC penalizes large models in attempt to select a model that performs well on the data without overfitting \cite{neath_bayesian_2012}. The BIC is formally defined as 

\begin{equation}
\text{BIC } = klog(n) - 2\text{log}(L(D|\theta)),
\end{equation}
where $L(D|\theta)$ is the likelihood of the data $D$ given the parameters $\theta$, $k$ is the number of parameters in $\theta$, and $n$ are the number of data points in $D$. Generally, the model with the smallest BIC is chosen. 

The model selection problem is well studied, and many different model selection criteria have been proposed. One generalization of the BIC is the Generalized Information Criterion (GIC) \cite{kim_consistent_nodate}. The GIC introduces a parameter $\lambda$ which is a general penalty term:

\begin{equation}
    GIC = k\lambda - 2log(L(\theta|D)).
\end{equation}
The BIC and the GIC are equivalent when $\lambda = log(n)$. Generally, $\lambda$ is used to change the penalty based on the size of the data set and its dimensionality. For higher-dimensional data, one choice of $\lambda$ that has been previously used is $\lambda = log(m)log(n)$, where $m$ is the number of features in the data \cite{kim_consistent_nodate}. To choose a model output the phylogeny-aware clustering algorithm, we use the GIC with $\lambda = log(m)log(n)$, which heavily penalizes larger models when the bulk data contains many samples.

\section{Evaluation metrics}\label{appendix:evaluation-metrics}

In this section, we describe the evaluation metrics used to score mutation trees, and mutation clusters.

\subsection{Perplexity}

We use the perplexity to evaluate a mutation tree based on its mutation frequency matrix $F$. Assuming we've sampled $I$ total unique trees $A = \{t_1,t_2,\dots,t_I\}$ with some probability distribution over this set of trees $0 \leq p(t_i) \leq 1$ with $\sum_{i=1}^I{p(t_i)} = 1$, we can compute the perplexity of this set of trees, $PP(A)$, as 

\begin{align}
    PP(A) &= 2^{\epsilon} \label{eq:log-perplexity} \\
    \epsilon &= - \frac{1}{nm}\sum_{j=1}^n\sum_{s=1}^mlog_2\left(\sum_{i=1}^I{p\left(X_{js}|F_{js}^{(i)}\right)p(t_i)}\right) \label{eq:perplexity-exponent}
\end{align}
where $F^{(i)}$ is the mutation frequency matrix for $i$-th tree, and  $p\left(X_{js}|F_{js}^{(i)}\right) = \text{Binom}(b_{js}|b_{js} + a_{js}, \omega_{js}F^{(i)}_{js})$. We choose the following definition for $p(t_i)$:

\begin{equation}\label{eq:likelihood-weighted}
p(t_i) = \frac{\text{log}P(D|B^{(i)},U^{(i)})}{\sum_{i^{\prime} = 1}^I\text{log}P(D|B^{(i^{\prime})},U^{(i^{\prime})})},
\end{equation}
where $B^{(i)}$ and $U^{(i)}$ are the binary genotype matrix and clonal proportion matrix for the $i$-th tree, respectively. The numerator and denominator in this equation are computed using Eq \ref{eq:approximate-data-likelihood-factor}. Using this definition means that Eq \ref{eq:log-perplexity} is a likelihood weighted average perplexity of the mutation frequency matrices $\{F^{(i)} : i = 1,\dots,I\}$ under a binomial sampling model. 

 We can measure the perplexity of a mutation frequency matrix reconstructed by a method in reference to a baseline perplexity using the ratio $\frac{2^{\epsilon_{\Omega}}}{2^{\epsilon_{\text{base}}}}$, where $\epsilon_{\Omega}$ is equivalent to Equation \ref{eq:perplexity-exponent} and $\epsilon_{\text{base}}$ is the exponent for the baseline perplexity being compared against. The perplexity can sometimes be very large; therefore, we choose to transform the perplexity and the perplexity ratio using a base 2 logarithm, resulting in the \textit{log perplexity} and \textit{log perplexity ratio}. The log perplexity ratio has also been called the \textit{VAF Reconstruction loss} \cite{wintersinger_reconstructing_2022}. We denote the log perplexity ratio as $\hat{\epsilon}$,  where $\hat{\epsilon} = log_2\left(\frac{2^{\epsilon_{\Omega}}}{2^{\epsilon_{\text{base}}}}\right) = \epsilon_{\Omega} - \epsilon_{\text{base}}$. Therefore, it is possible that $\hat{\epsilon}$ is negative showing that the mutation frequency matrix $F$ for the tree(s) reconstructed by a method fit the VAF data better than the baseline mutation frequency matrix. With simulated data, the baseline is the ground-truth mutation frequency matrix $F^{\text{(true)}}$ used to generate the simulated VAF data. With real data, the baseline is a \textit{maximum a posteriori} (MAP) mutation frequency matrix $F^{(\text{MAP})}$ fit to the expert derived tree. The MAP mutation frequency matrices fit to the expert-derived trees were calculated using a gradient based method called \textit{resilient backpropagation} (rprop) \cite{wintersinger_reconstructing_2022}, which is described in the next section. 

\subsection{Computing the \textit{maximum a posteriori} mutation frequency matrix for expert-derived trees}\label{appendix:rprop}

A bulk DNA clone tree derived by experts may not have a mutation frequency matrix that adheres to a perfect phylogeny. However, we can obtain a very close approximation of its mutation frequency matrix by fitting a maximum a posteriori (MAP) estimate of $F$ after observing the expert-derived tree. Although Orchard uses the projection algorithm \cite{jia_efficient_2018} to fit the mutation frequency matrix $F$, this algorithm optimizes a Gaussian approximation to the binomial likelihood, so it is not guaranteed to find the maximum likelihood estimate. However, a very close approximation to the maximum likelihood of
$F$ can be obtained using gradient-based methods. Unfortunately, these methods take a long time to converge, making them impractical for large reconstruction problems or in situations where many $F$ matrices need to be estimated. However, gradient-based methods are quite useful for fitting a MAP estimate of $F$ given a clone tree derived by experts. One such gradient-based method used for this task in prior research is rprop, or \textit{resilient backpropagation} \cite{wintersinger_reconstructing_2022}. We chose to use rprop to find the MAP mutation frequency matrix, $F^{(\text{MAP})}$, for the B-ALL expert-derived clone trees. We ran rprop for 30,000 epochs on each B-ALL data set to estimate $F^{(\text{MAP})}$ for each expert-derived clone tree.

\subsection{Relationship reconstruction loss}

Given a mutation tree $t = \{V, E, M\}$ and an ordered pair of nodes $(u,v) \in V$, there are 3 possible pairwise evolutionary relationships the ordered pair $(u,v)$ can have:

\begin{definition}[Pairwise Evolutionary Relationships]\label{supdef:pairwise-rel}
\begin{description}
\item
    \item \textbf{Ancestral}: u is an ancestor of v, i.e., v contains the mutation(s) associated with u, but v contains one or more mutations not present in u. 
    \item \textbf{Descendant}: v is an ancestor of u, i.e., the same as above but u and v are switched. 
    \item \textbf{Branched}: u and v share some common ancestor, but neither u nor v are ancestral to each other.
\end{description}
\end{definition}
We denote the 3 possible pairwise evolutionary relationships from Definition \ref{supdef:pairwise-rel} for the ordered pair of nodes $(u,v)$ as one of the following:

\begin{align*}
    R_{uv} &= \text{ ancestral } \\
    R_{uv} &= \text{ descendant } \\
    R_{uv} &= \text{ branched }, 
\end{align*}
where $R_{uv}$ denotes the evolutionary relationship between the ordered pair $(u,v)$.
Note that $R_{uv}$ can be derived from the binary genotype matrix $B$ associated with a tree $t$ as follows:
\begin{equation*}
  R_{uv} =
  \begin{cases}
    \text{ ancestral } & \text{if}\ B_{uv} = 1, B_{vu} = 0\\
    \text{ descendant } & \text{if}\ B_{uv} = 0, B_{vu} = 1\\
    \text{ branched } & \text{if}\ B_{uv} = B_{vu} = 0
  \end{cases}
\end{equation*}
and if $B_{uv} = B_{vu} = 1$ then $B$ is not consistent with any tree, as this implies a cycle.

The \textit{relationship reconstruction loss} measures how well the pairwise relationships between mutations in some proposed tree $t$ match the pairwise relationships in a set of $N$ ground truth trees:

\begin{equation*}
A^{(\text{true})} = \left\{t^{(\text{true})}_1, t^{(\text{true})}_2, \dots, t^{(\text{true})}_N\right\}.
\end{equation*}
We primarily use the relationship reconstruction loss when evaluating reconstructions for simulated mutation tree reconstruction problems. We generated simulated trees and VAF data using the Pearsim software (https://github.com/morrislab/pearsim). The Pearsim software starts with a ground truth mutation frequency matrix $F^{(\text{true})}$ and enumerates all possible trees that fit $F^{(\text{true})}$ without violating the ISA. This process results in a set $A^{(\text{true})}$. Since each pair of nodes $(u,v)$ in a tree $t$ can have one of the three mutually exclusive pairwise relationships (see Definition \ref{supdef:pairwise-rel}), we define the probability that $(u,v)$ has a particular pairwise evolutionary relationship in $t$ as 

\begin{equation*}
    p(R_{uv} = e|t) = \begin{cases}
        1 & \text{ iff } R_{uv} = e \text{ in } t \\
        0 & \text{ otherwise},
    \end{cases}
\end{equation*}
where $e \in \{\text{ancestral}, \text{descendant}, \text{branched}\}$ as defined in Definition \ref{supdef:pairwise-rel}, and $R_{uv}$ is the pairwise evolutionary relationship between nodes $u$ and $v$ defined by the tree $t$.

Let $A = \left\{t_1, t_2, \dots, t_I\right\}$ be the set of $I$ unique trees reconstructed by a MSPP method. We can compute the probability that a particular evolutionary relationship $e$ occurs between nodes $u$ and $v$ in this set as 

\begin{equation}
p(R_{uv} = e|A) = \sum_{i=1}^I{p(R_{uv} = e|t_i})p(t_i).    
\end{equation}
We use the same definition for $p(t_i)$ from Eq \ref{eq:likelihood-weighted}. We can compute the same probability with the set of $N$ true trees, only we use a uniform prior $p(t^{(\text{true})}) = \frac{1}{N}$. We denote the posterior distribution over the pairwise relationships in our set of $I$ proposed trees as $p(R_{uv}|A)$, while we denote the posterior distribution over the pairwise relationships in our set of $N$ true tree as $p(\Tilde{R}_{uv}|A^{(\text{true})})$. We measure the difference between these two distributions using the Jensen-Shannon divergence (JSD) while normalizing over the total number of pairs of mutations in the tree 

\begin{equation}
    \epsilon_{R} = \frac{2}{n(n+1)}\sum_{u,v}{JSD(R_{uv} || \Tilde{R}_{uv})}
\end{equation}

\subsection{Adjusted Rand Index}

The Rand Index is used to compare two clusterings based on the pairs of elements that co-occur (or do not co-occur) in the same cluster. The adjusted Rand Index is a normalized version of the Rand Index. The Rand Index is a value between 0 and 1, while the adjusted Rand Index is a value between -1 and 1, where a value below 0 is obtained if the difference between the two clusterings is worse than what would be expected if the clusterings were randomly generated. For a bulk data set containing $n$ mutations, there are $\binom{n}{2}$ possible pairs of mutations, denoted by the set $A$. The set $P$ represents the pairs of mutations that co-occur in the same clusters as determined by a mutation clustering algorithm. The set $T$ represents all pairs of mutations that co-occur in the baseline clusters, such as those from ground truth or expert-derived data. We use these sets to define the following:

\begin{itemize}
    \item $TP = |P \cap T|$ is the number of \textit{true positives}.
    \item $FP = |P \setminus T|$ is the number of \textit{false positives}.
    \item $FN = |T \setminus P|$ is the number of \textit{false negatives}.
    \item $TN = |(A \setminus P) \cap (A \setminus T)|$ is the number of \textit{true negatives}.
\end{itemize} 
 The Rand Index can be defined in terms of the number of pairs of mutations that are classified as true positives (TP), false positives (FP), false negatives (FN), and true negatives (TN). We compute the Rand Index as 

\begin{equation}
    RI = \frac{TP + TN}{TP + FP + FN + TN},
\end{equation}
and we can compute the adjusted Rand Index as 

\begin{equation}
    ARI = \frac{2(TP \times TN - FN \times FP)}{((TP + FN) \times (FN + TN) + (TP + FP) \times (FP + TN))}.
\end{equation}

\section{Additional Experiments}

\subsection{Benchmarking on 576 simulated clone tree data sets}\label{appendix:ptsims}

We also evaluated Orchard, Pairtree, and CALDER on 576 simulated clone tree data sets originally from \cite{wintersinger_reconstructing_2022}. These clone tree data sets were simulated using the \textit{Pearsim} software (https://github.com/morrislab/pearsim). Each of the 576 simulated clone tree data sets had varying numbers of subclones (3, 10, 30, 100), average mutations per subclone (10, 20, 100), cancer samples (1, 3, 10, 30, 100), and sequencing depths (50x, 200x, 1000x).

\begin{figure}[h]
\includegraphics[width=\textwidth]{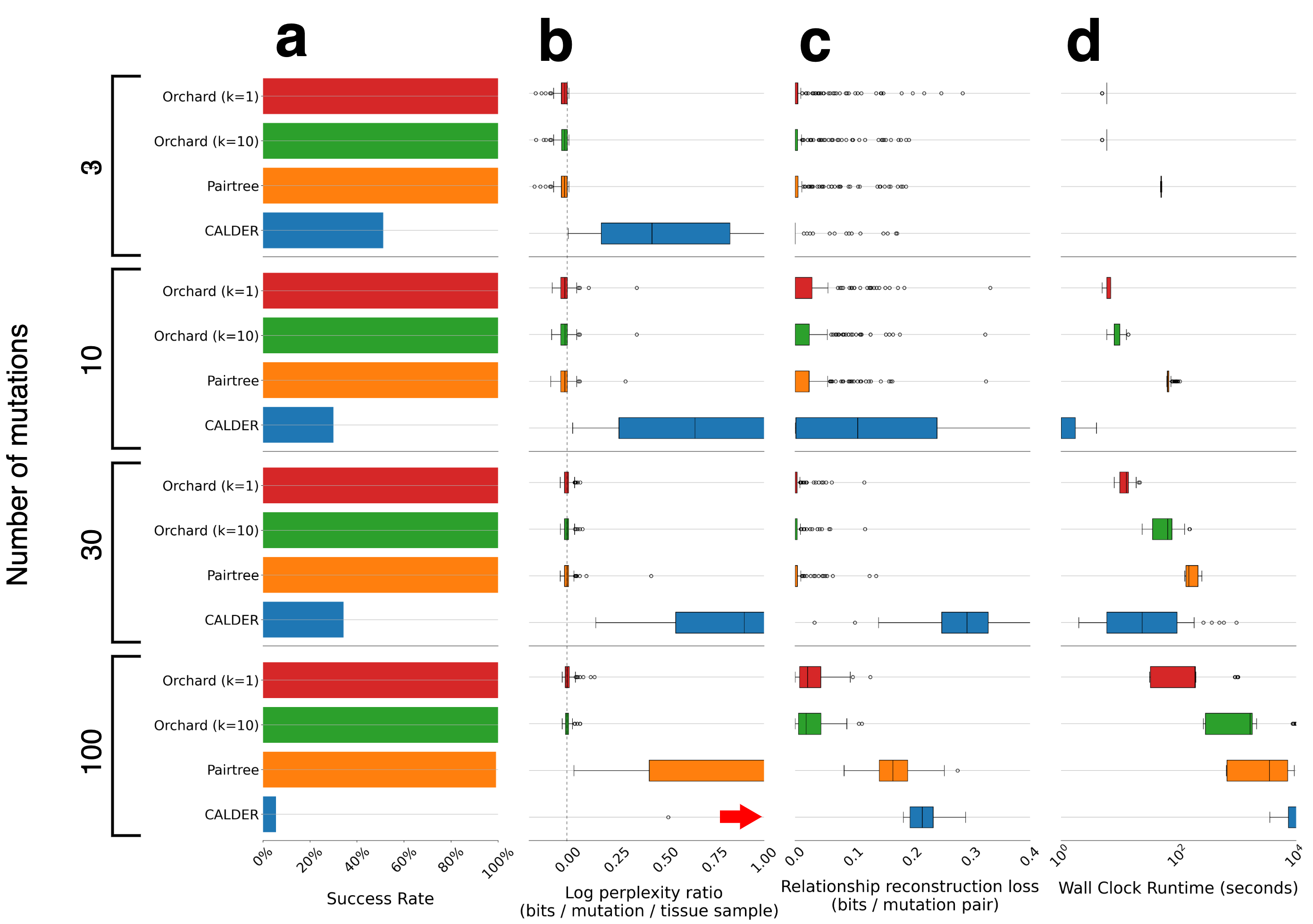}
\caption{Benchmark performance on 576 simulated data sets from \cite{wintersinger_reconstructing_2022}. The simulation results are grouped by the number of subclones (rows), and these groups are referred to as a \textit{problem size}. \textbf{A.} Bar plots show the success rate for each method on a problem size. A method is successful on a reconstruction problem if it produces at least one valid tree. The distributions in b-d only reflect data sets where a method was successful. \textbf{B.} Box plots show the distribution of log perplexity ratio for each method on a problem size. Log perplexity ratio is reported relative to the true mutation frequency matrix $F^{(\text{true})}$ used to generate the VAF data, and therefore can be negative. A red arrow means the results for the method on a problem size occur beyond the x-axis limit. \textbf{C.} The distribution of relationship reconstruction loss for each method on a problem size. \textbf{D.} The distributions of wall clock run time in seconds.}
\label{fig:ptsims-results}
\end{figure}

Figure \ref{fig:ptsims-results} shows the performance of Orchard, CALDER, and Pairtree on the 576 simulated clone tree data sets. These results generally reiterate the findings described in Section \ref{sec:simulated-results}. Orchard and Pairtree were able to succeed on all 576 simulated clone tree data sets. CALDER succeeded on 51\% (92/180) of 3-clone data sets, 30\% (54/180) of 10-clone data sets, 34\% (37/108) of 30-clone data sets, and 6\% (6/108) of 100-clone data sets. On the 3-clone, 10-clone, and 30-clone data sets, Orchard and Pairtree had similar performance for log perplexity ratio, and relationship reconstruction loss. Orchard in both of its configurations ($k=1$ and $k=10$) had faster run times than Pairtree on all data sets with 30 or fewer clones. On datasets with 100 clones, CALDER outperformed Pairtree in both log perplexity ratio and relationship reconstruction loss on the subset of datasets where it was successful. Meanwhile, Orchard far outperformed both CALDER and Pairtree on all 100-clone data sets for both metrics. Orchard $(k=1)$ was between 5-100x faster than both CALDER and Pairtree on 100-clone data sets, and Orchard $(k=10)$ had similar run times to Pairtree on the 100-clone data sets. 

One notable result in Figure \ref{fig:ptsims-results} is that CALDER has more successes on the 30-clone datasets compared to the 10-clone datasets. We believe this phenomenon is primarily due to CALDER's optimization routine, which at times discards data if certain constraints are violated. Although the Gurobi optimizer (https://www.gurobi.com/) CALDER uses to solve its mixed integer linear program (MILP) problem formulation is deterministic, the solution path it takes is machine dependent since libraries and other utilities may vary between different machines. Therefore, it's possible for CALDER to discard different data during separate runs on separate machines, since the solution path may be different. This may explain why we get more failures on 10-clone data sets compared to 30-clone data sets, and more generally, why the results shown here are different than those shown in \cite{wintersinger_reconstructing_2022}. 

\subsection{Comparing Orchard with greedy sampling on 90 simulated cancers}

\begin{figure}[ht]
\begin{center}
\includegraphics[scale=0.09]{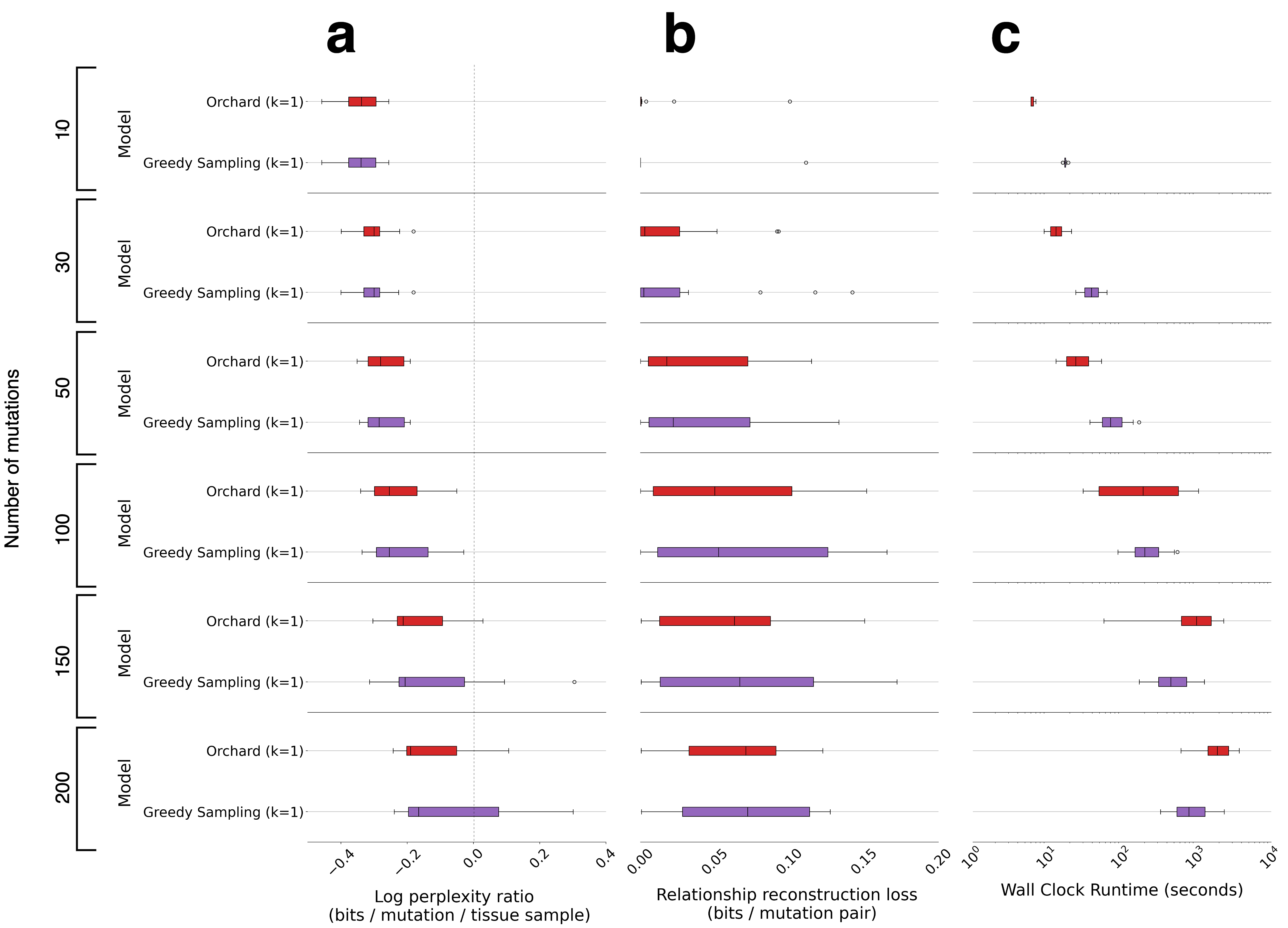}
\end{center}
\caption{Comparison of Orchard vs. Greedy sampling on the 90 simulated mutation tree data sets. The simulation results are grouped by the number of mutations (rows), and these groups are referred to as a \textit{problem size}. These results show that Orchard generally produces better reconstructions compared to Greedy sampling, with the caveat of slightly longer run times. \textbf{a. } The distribution of log perplexity ratio for each method on a problem size. Log perplexity ratio is reported relative to the true cellular prevalence matrix $F^{(\text{true})}$ used to generate the data, and therefore can be negative. \textbf{b.} The distribution of relationship reconstruction loss for each method on a problem size. \textbf{c.} The distributions of wall clock run time in seconds.}
\label{supfig:greedy-sampling}
\end{figure}

An alternative approach to sampling from $Q^{\pi}(B|D)$ is ``greedy" sampling, as described in Section \ref{sec:greedy-sampling}. Figure \ref{supfig:greedy-sampling} compares Orchard vs. Greedy sampling across the 90 simulated mutation tree data sets. The results in Figure \ref{supfig:greedy-sampling}a-b indicate that Orchard's sampling approach reduces the variance of the log perplexity ratio and improves relationship reconstruction loss, resulting in slightly better reconstructions compared to a greedy search strategy. Figure \ref{supfig:greedy-sampling}c demonstrates that Greedy sampling generally outperforms Orchard in terms of speed for larger mutation tree reconstruction problems (100+ mutations). These findings suggest that while Greedy sampling's runtime scales more favorably for larger problems, it may sacrifice reconstruction quality.

\subsection{Comparing Orchard using the \texorpdfstring{$\widehat{F}-$}sum vs. randomized mutation order}

The accuracy of $Q^{\pi}(B|D)$ depends on the mutation order $\pi$. In Appendix \ref{appendix:mutation-ordering}, we introduced the $\widehat{F}-$sum mutation order and discussed why it can improve the accuracy of $Q^{\pi}(B|D)$. Here, we run Orchard with both the $\widehat{F}-$sum ordering and a randomized order, and evaluate their impact on Orchard's ability to sample from $Q^{\pi}(B|D)$.

\begin{figure}[ht]
\begin{center}
\includegraphics[scale=0.1]{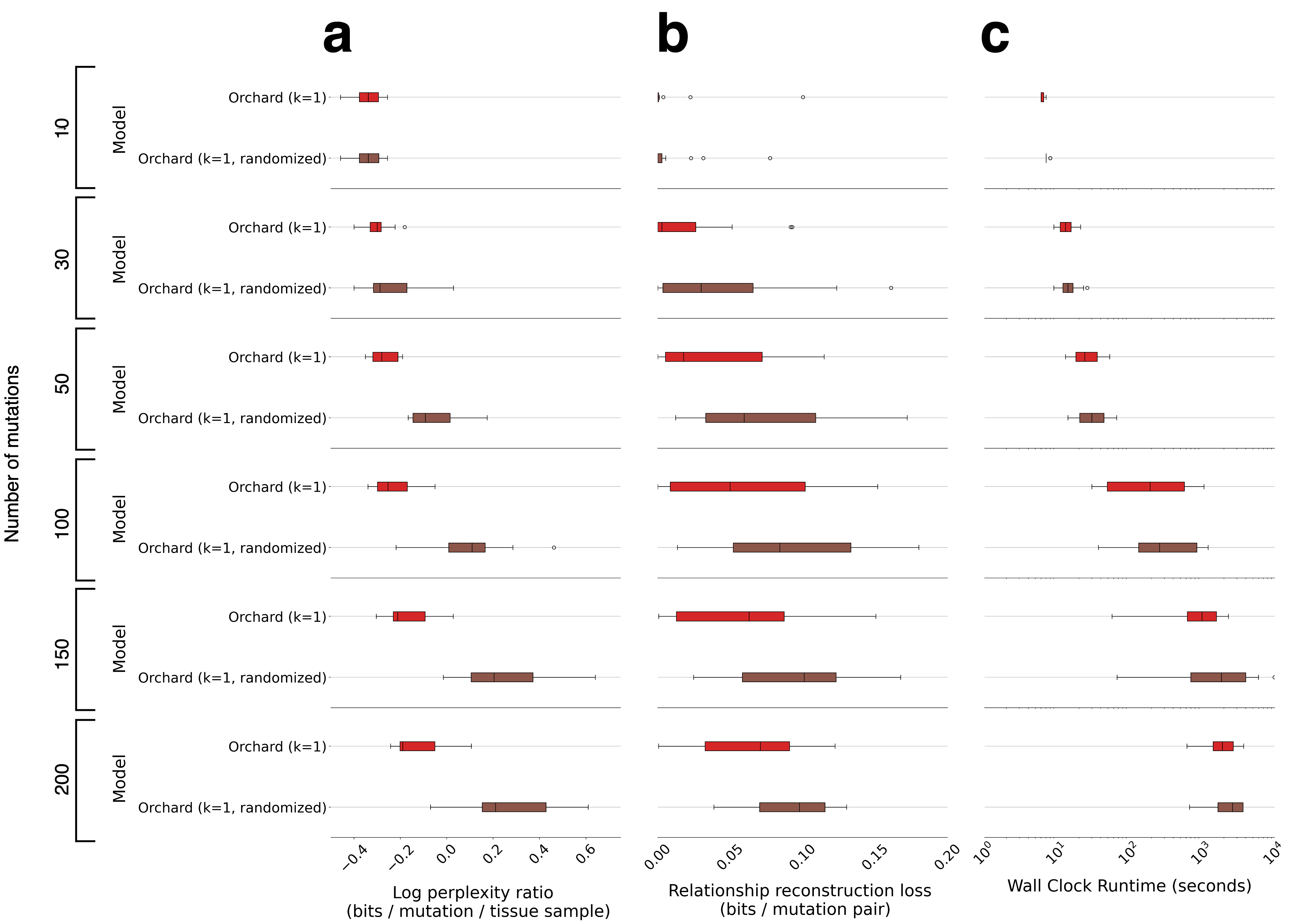}
\end{center}
\caption{Comparison of Orchard's performance with a randomized mutation order versus the $\widehat{F}-$sum order on 90 simulated mutation tree data sets. The simulation results are grouped by the number of mutations (rows), and these groups are referred to as a \textit{problem size}. These results show that using a randomized order can result in slightly worse reconstructions compared to using the $\widehat{F}$ sum order described in Appendix \ref{appendix:mutation-ordering}. \textbf{a.} The distribution of log perplexity ratio for each method on a problem size. Log perplexity ratio is reported relative to the true cellular prevalence matrix $F^{(\text{true})}$ used to generate the VAF data, and therefore can be negative. \textbf{b.} The distribution of relationship reconstruction loss for each method on a problem size. \textbf{c.} The distributions of wall clock run time in seconds.}
\label{fig:mutsim-random}
\end{figure}

Orchard was run with the parameters $k=1$ and $f=20$. We generated randomized mutation orders by randomly permuting $\pi$. We denote Orchard run with the randomized order as ``Orchard (k=1, randomized)". In Figure \ref{fig:mutsim-random}, we compare Orchard's reconstructions using the $\widehat{F}-$sum and randomized mutation orders on the 90 simulated mutation tree data sets. As shown in Figure \ref{fig:mutsim-random}a-b, randomizing the mutation order results in a slightly higher log perplexity ratio and relationship reconstruction loss compared to the $\widehat{F}-$sum order, indicating worse reconstructions. Additionally, Figure \ref{fig:mutsim-random}c shows that randomizing the mutation order leads to slightly longer run times. We believe this increase in run times and the worse reconstructions are related; poor mutation placements force the projection algorithm \cite{jia_efficient_2018} to adjust more values in $U$ to adhere to the perfect phylogeny constraints, resulting in increased run times.

Although randomizing the mutation order results in slightly worse reconstructions, it's evident that $Q^{\pi}(B|D)$ remains an accurate approximation regardless of the setting of $\pi$. These results also support that the heuristic $H(t^{(\ell)},f)$ described in Appendix \ref{appendix:placing-mutations} is robust to any mutation order.

\subsection{Evaluating Orchard on 1000-node mutation trees}\label{appendix:1000k-mutation-trees}

To evaluate Orchard's ability to reconstruct extremely large mutation trees, we selected four datasets from the 576 simulated clone trees used in Appendix \ref{appendix:ptsims}, each containing 1,000 mutations. These data sets had 1,000 mutations, 10 samples, and a read depth of 50x.

Orchard completed each reconstruction in approximately $26 \pm 0.5$ hours. These reconstructions closely matched the data fit of the ground truth mutation frequency matrix $F^{(\text{true})}$ used to generate the simulated VAF data, achieving an average log perplexity ratio of 0.025 across the four datasets. However, the pairwise relationships recovered by these trees were somewhat crude, with an average relationship reconstruction loss of 0.33. These results are expected, considering that a read depth of 50x may not be adequate for the precise resolution of all pairwise relationships. However, they demonstrate that Orchard can still identify trees with a good fit to the data despite this limitation.



\subsection{Mutation tree reconstructions for the 14 B-ALL data sets}\label{appendix:b-all-mutation-trees}

\begin{figure}
\begin{center}
\includegraphics[scale=0.28]{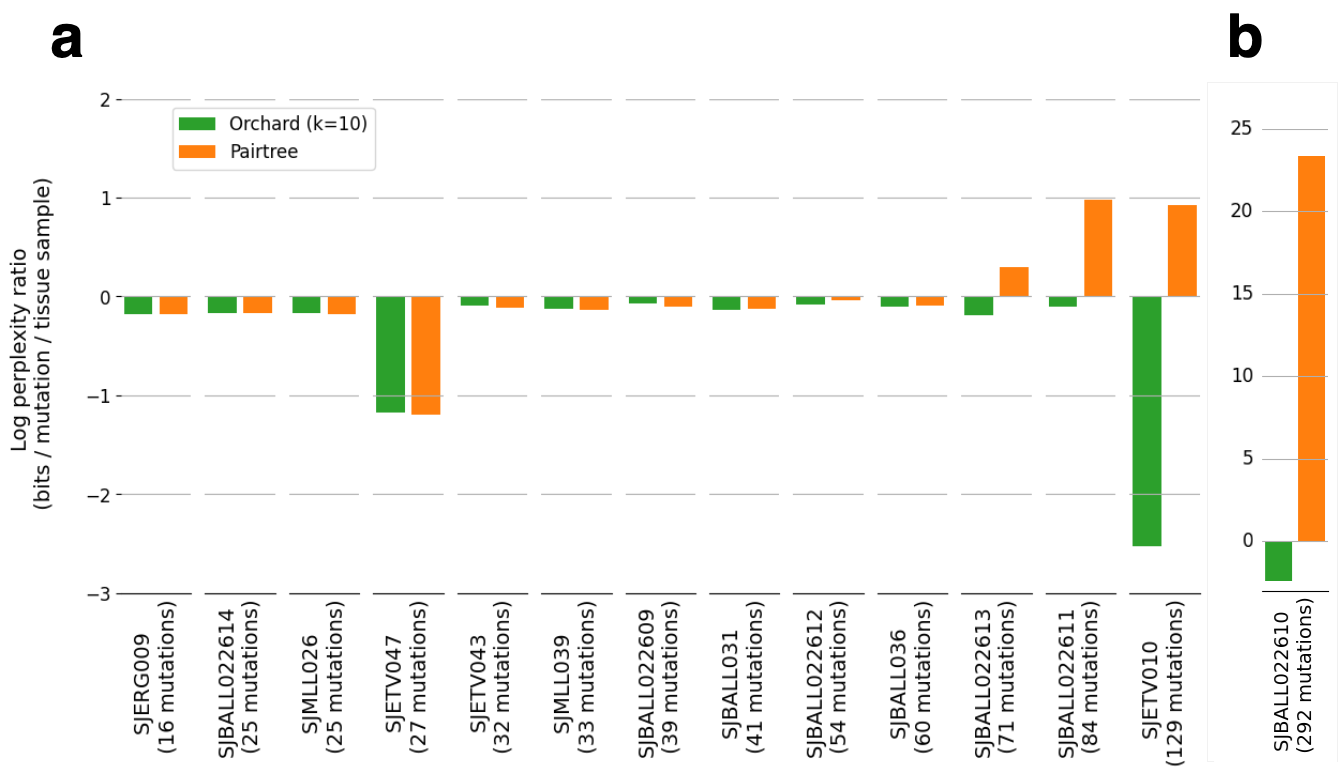}
\caption{Evaluation of mutation tree reconstructions for the 14 B-ALL data sets. Log perplexity ratio is reported relative to the maximum a posteriori (MAP) mutation frequencies $F^{(\text{MAP})}$ for the expert derived clone trees. \textbf{a-b.} Bar plots show the log perplexity ratio for the mutation trees reconstructed by Orchard ($k=10$) and Pairtree using the mutation data for each B-ALL data set. Part \textbf{b} has a larger y-axis range to accommodate Pairtree's large log perplexity ratio on the 292 node data set.}
\label{fig:ball-mutation-trees}
\end{center}
\end{figure}

In this section, we evaluate mutation tree reconstructions for the B-ALL data. These 14 data sets contain between 16 and 292 mutations, with a median of 40, resulting in very large mutation trees. Due to CALDER's high failure rate on data sets with more than 30 mutations (see Figure \ref{fig:90mutsim_compare}), it was excluded from this analysis. As we do not have expert-derived mutation trees for the 14 B-ALL data sets, we use the MAP mutation frequency matrix $F^{(\text{MAP})}$ for each data set's expert-derived clone tree as a baseline. Figure \ref{fig:ball-mutation-trees}a-b shows the log perplexity ratio for Orchard ($k=10$) and Pairtree on each B-ALL data set. It illustrates that while Orchard matches the performance of Pairtree on reconstruction problems with fewer than 50 mutations, it significantly outperforms Pairtree on problems with more than 50 mutations. Notably, on mutation tree reconstruction problems with 129 mutations (SJETV010, Figure \ref{fig:ball-mutation-trees}a) and 292 mutations (SJBALL022610, Figure \ref{fig:ball-mutation-trees}b), Orchard beats Pairtree by more than $3$ and $25$ bits, respectively. The results in Figure \ref{fig:ball-mutation-trees}a-b also illustrates that the mutation trees reconstructed by Orchard consistently exhibit better agreement with the VAF data compared to the expert-derived clone trees. For three particular cases (SJVET047, SJETV010, SJBALL0222610), the mutation trees reconstructed by Orchard have a relative decrease in perplexity that is greater than 1 bit, suggesting the possibility that the expert-derived mutation clusters for these patients are incorrectly grouping distinct subclones together.

\subsection{Evaluating the phylogeny-aware clustering algorithm on 14 B-ALL data sets}\label{appendix:ball-pac}

In this section, we compare the mutation clusters inferred by the ``phylogeny-aware" clustering algorithm to those inferred by VAF based clustering methods on the 14 B-ALL data sets. The phylogeny-aware clustering algorithm was provided the best mutation tree reconstructed by Orchard and Pairtree for each B-ALL data set, corresponding to those in Section \ref{appendix:b-all-mutation-trees}. We then use three state-of-the-art VAF-based mutation clustering methods on each data set: PyClone-VI \cite{gillis_pyclone-vi_2020}, VIBER \cite{caravagna_subclonal_2020}, and SciClone \cite{miller_sciclone_2014}. We compare the subclones output by each method to the expert-defined clusters using the \textit{Adjusted Rand Index} (ARI). The ARI is a measurement of the percentage of correctly co-clustered mutations adjusted for randomness \cite{santos_use_2009}.

\begin{figure}
\centering
\includegraphics[scale=0.4]{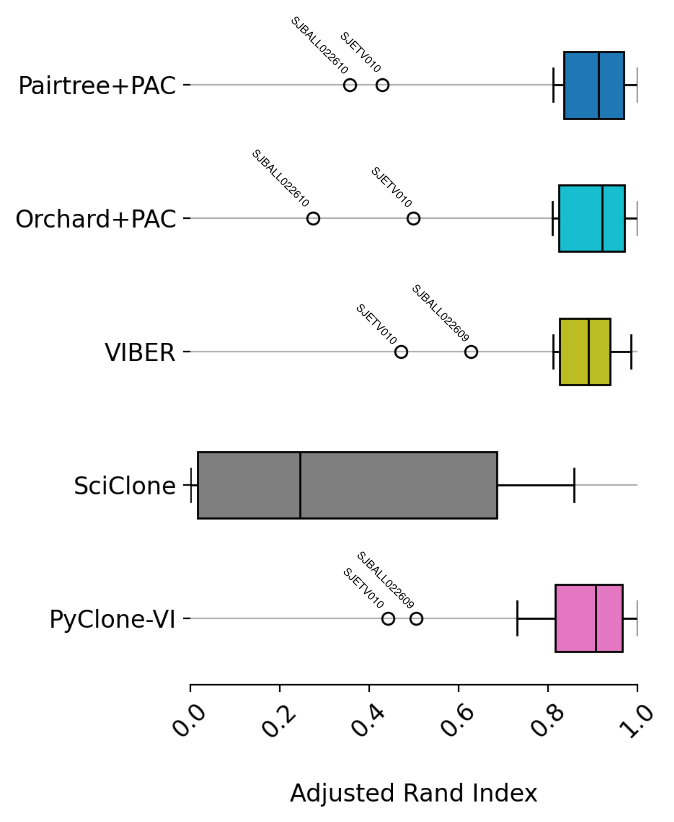}
\caption{Box plots of the Adjusted Rand Index (ARI) comparing the "phylogeny-aware" clustering method with state-of-the-art VAF-based mutation clustering methods across 14 B-ALL data sets. Adjusted Rand Index is measured in reference to the expert-derived mutation clusters for the 14 B-ALLs. Outliers are labeled by their unique ID.}
\label{supfig:ball-clustering}
\end{figure}

Figure \ref{supfig:ball-clustering} compares the distributions of the Adjusted Rand Index for each method across the 14 B-ALL data sets. We use ``Orchard+PAC" and ``Pairtree+PAC" to refer to the phylogeny-aware clustering method applied to mutation trees constructed by Orchard and Pairtree, respectively. Mutation clusters inferred by ``Orchard+PAC" were most similar to the expert-derived mutation clusters for the majority of data sets. The breakdown of each method's performance on the individual data sets are shown in Table \ref{table:ball-clustering-results}. While the phylogeny-aware clustering method matched or outperformed all other clustering methods across the majority of B-ALL data sets, it fell short in recovering mutation clusters competitive with state-of-the-art VAF-based clustering methods specifically for SJBALL022610. In fact, for SJBALL022610, the clusters inferred by the phylogeny-aware method using Orchard's tree were poorer than those inferred using Pairtree's tree, despite Orchard's reconstructed tree being significantly better than Pairtree's (see Figure \ref{fig:ball-mutation-trees}). This could be due to the algorithm's clustering model being too simplistic for handling such a large data set with 292 mutations detected across 27 samples. It employs a non-probabilistic merging strategy to identify clones in mutation trees, with additional phylogenetic constraints that dictate when nodes can be merged. It could also be the case that the trees reconstructed by both Pairtree and Orchard for SJBALL022610 are too crude to accurately recover clones.

\subsection{Comparing expert-derived trees to those inferred using ``phylogeny-aware" clustering}\label{appendix:tree-based-analysis}

In this section, we use the mutation clusters from Section \ref{appendix:phylogeny-aware-clustering} to construct clone trees. Note that trees with more clones tend to have a better data fit because the clonal frequencies can align more closely to fewer mutations. It's important to consider this when comparing results between methods that output clone trees of different sizes. To evaluate our new clonal tree reconstruction approach against the classical approach, which we introduced in Section \ref{sec:real-results}, we reconstructed clonal trees for the 14 B-ALL data set using the following methods:

\begin{figure}[H]
\begin{center}
\includegraphics[scale=0.52]{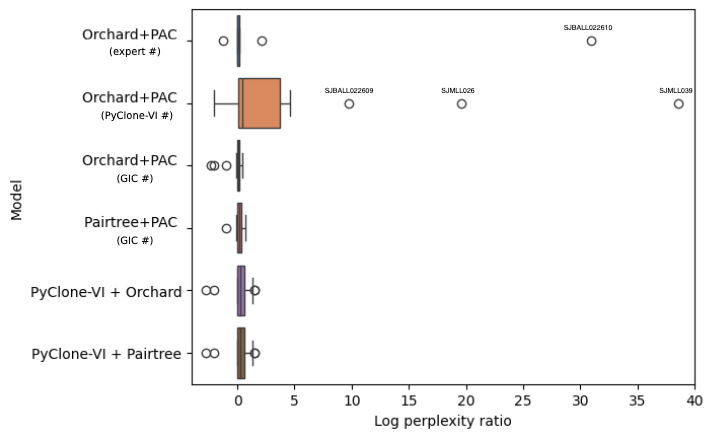}
\end{center}
\caption{The box plots illustrate the distribution of log perplexity ratios for trees inferred by phylogeny-aware clustering compared to clone trees reconstructed using clusters output by PyClone-VI across 14 B-ALL data sets from \cite{dobson_relapse-fated_2020}. The log perplexity ratio is calculated relative to $F^{(\text{MAP})}$ fitted to the expert-derived tree for each data set. Overall, the phylogeny-aware clustering algorithm using the GIC for model selection, ``Orchard+PAC (GIC \#)", performed the best.}
\label{fig:tree-based-analysis}
\end{figure}

\begin{itemize}
    \item ``Orchard+PAC (expert \#)": phylogeny-aware clustering applied to Orchard's best reconstructed tree; manually selected clone tree with the same number of clones as the expert-derived tree.
    \item ``Orchard+PAC (PyClone-VI \#)": phylogeny-aware clustering applied to Orchard's best reconstructed tree; manually selected clone tree with the same number of clones as recovered by PyClone-VI.
    \item ``Orchard+PAC (GIC \#)": phylogeny-aware clustering applied to Orchard's best reconstructed tree; selected clone tree using GIC.
    \item ``Pairtree+PAC (GIC \#)": phylogeny-aware clustering applied to Pairtree's best reconstructed tree; selected clone tree using GIC.
    \item ``PyClone-VI+Orchard": Clone tree reconstructed by Orchard using the clones recovered by PyClone-VI.
    \item ``PyClone-VI+Pairtree": Clone tree reconstructed by Pairtree using the clones recovered by PyClone-VI.
\end{itemize}
To ensure a fair comparison among trees, we manually selected trees of the same size from the phylogeny-aware clustering output to compare against the expert-derived trees and the clone trees reconstructed using PyClone-VI's clusters. We only used PyClone-VI in this comparison because it performed the best among all VAF-based mutation clustering methods in our experiments.

\begin{figure}[H]
\begin{center}
\includegraphics[scale=0.55]{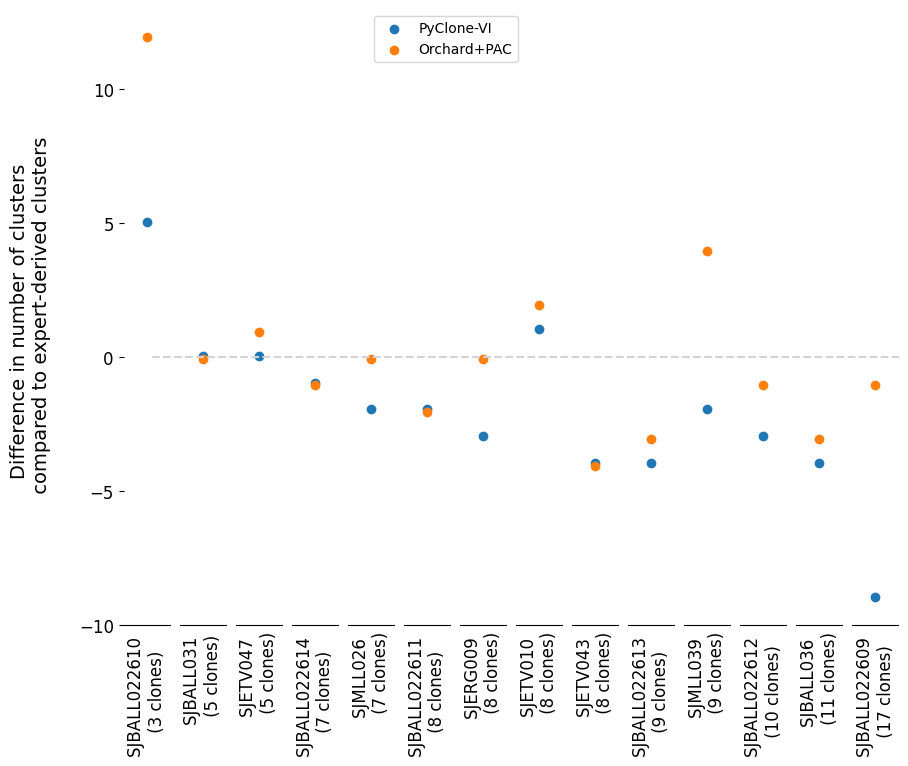}
\end{center}
\caption{Difference in the number of clones inferred by ``Orchard+PAC (GIC \#)" versus PyClone-VI \cite{gillis_pyclone-vi_2020} for the 14 B-ALL data sets. The number of clones inferred by each method is relative to the number of expert-derived clones. Each column shows the results for a single B-ALL data set. The horizontal gray dotted line represents no difference in the number of clusters inferred by a method compared to the expert-derived number. A point that lies far above or below this gray dotted line means that a method inferred a number of clone that's very different from the expert-derived number.}
\label{fig:ball-cluster-nums}
\end{figure}

Figure \ref{fig:tree-based-analysis} shows the distributions of log perplexity ratios for the clone trees reconstructed by each method on the 14 B-ALL data sets. The log perplexity ratio is computed in relation to $F^{(\text{MAP})}$ for each of the expert-derived clone trees. Overall, ``Orchard+PAC (GIC \#)" generally found the best clone trees of any method. It's important to note that this method greatly overestimated the number of clones for SJBALL022610 compared to the expert-derived count. However, for all other data sets, the estimated number of clones was, on average, closer to the expert-derived count compared to PyClone-VI (see Figure \ref{fig:ball-cluster-nums}). One notable finding is that "Orchard+PAC (PyClone-VI \#)" yielded several poor reconstructions when selecting the clone tree with the same number of clones as PyClone-VI's output. There are three B-ALL data sets identified as outliers for this method in Figure \ref{fig:tree-based-analysis} that range from about +10-40 bits above the baseline. We believe this finding is largely due to PyClone-VI inferring far fewer clones on the B-ALL data sets compared to the experts. For 10/14 B-ALLs PyClone-VI infers on average 3.4 fewer clones per data set compared to the expert-derived counts. In contrast to "Orchard+PAC (PyClone-VI \#)", "Orchard+PAC (expert \#)" performs very well, with the only dramatic outlier being SJBALL022610. These findings suggest that, for all B-ALLs besides SJBALL022610, the phylogeny-aware clustering algorithm aligns more closely with the number of clones derived by experts compared to the numbers output by PyClone-VI. This is reinforced by the results for ``Orchard+PAC (GIC \#)", which selects a clone tree that, on average, differs by $\pm 2$ clones compared to the expert count, whereas PyClone-VI differs by $\pm 2.5$ clones (see Figure \ref{fig:ball-cluster-nums}). Overall, this experiment supports that our proposed method of inferring clones from a mutation tree can yield better clone trees compared to the classical approach of reconstructing clone trees using clusters output by a VAF-based clustering method.

\section{Tables}

\begin{table}[H]
\centering
\begin{tabular}{|*{5}{c|}}  
\hline
 & Orchard (k=1) & Orchard (k=10) & Pairtree & CALDER \\ \hline
 SJBALL022609 & \textbf{0.016674}	 & 0.016822 & 0.016706 & 2.617256	 \\ \hline
SJBALL022610 & \textbf{0.000924}	& *0.000924 & *0.000924 & 1.648541 \\ \hline
SJBALL022611 & \textbf{0.000777} & 0.000800 & 0.000791 & 0.444308 \\ \hline
SJBALL022612 & \textbf{0.003345} & 0.003367 & 0.003348 & - \\ \hline
SJBALL022613 & \textbf{0.010827} & *0.010827 & *0.010827 & 1.052539 \\ \hline
SJBALL022614 & \textbf{0.006357} & *0.006357 & *0.006357 & 1.350474 \\ \hline
SJBALL031 & -0.000357 & -0.000357 & \textbf{-0.000386} & 0.604383 \\ \hline
SJBALL036 & \textbf{0.034083} & 0.054480 & *0.034083 & 0.905851\\ \hline
SJERG009 & \textbf{0.009929} & *0.009929 & *0.009929 & - \\ \hline
SJETV010 & 0.093638	& \textbf{0.083516} & *0.083516 & 1.556607 \\ \hline
SJETV043 & \textbf{0.009944} & 0.010110 & 0.010023 & 1.578067 \\ \hline
SJETV047 & \textbf{0.021099} & *0.021099 & *0.021099 & 1.538149 \\ \hline
SJMLL026 & 0.002358 & 0.002553 & \textbf{0.002446} & 1.201738 \\ \hline
SJMLL039 & *0.012682 & *0.012682 & \textbf{0.012682} & 1.438552 \\ \hline

\end{tabular}
\caption{Log perplexity ratio for each method on the 14 B-ALL clone tree reconstruction problems. Bold designates lowest (best) log perplexity ratio. A ``$*$'' represents a difference of less than 1e-6 compared to the best performing method. For display purposes, we do not show the full precision of the scores, so for some of the data sets, multiple methods may appear to have equal scores. Orchard (k=1) has a lower log perplexity ratio compared to Pairtree on 10/14 data sets.}
\label{table:ball-clone-tree-results}
\end{table}

\begin{table}[H]
\centering
\begin{tabular}{|*{6}{c|}}  
\hline
 & Orchard+PAC & Pairtree+PAC & PyClone-VI & VIBER & SciClone \\ \hline
 SJBALL022609 & 0.806 & \textbf{0.960} & 0.503 & 0.626 & 0.121 \\ \hline
SJBALL022610 & 0.273 & 0.356 & 0.730 & \textbf{0.984} & 0.019 \\ \hline
SJBALL022611 & \textbf{0.966} & 0.825 & \textbf{0.966} & \textbf{0.966} & 0.857 \\ \hline
SJBALL022612 & \textbf{0.893} & 0.891 & 0.891 & 0.888 & 0.742 \\ \hline
SJBALL022613 & 0.919 & 0.904 & \textbf{0.920} & 0.914 & 0.696 \\ \hline
SJBALL022614 & \textbf{0.971} & \textbf{0.971} & \textbf{0.971} & 0.855 & 0.709 \\ \hline
SJBALL031 & \textbf{1.0} & \textbf{1.0} & \textbf{1.0} & 0.946 & 0.321	 \\ \hline
SJBALL036 & \textbf{0.860} & \textbf{0.860}	 & 0.829 & 0.815 & 0.167 \\ \hline
SJERG009 & \textbf{0.920} & \textbf{0.920} & 0.875 & \textbf{0.875} & 0.0 \\ \hline
SJETV010 & 0.497 & 0.428 & 0.440 & 0.469 & \textbf{0.647} \\ \hline
SJETV043 & \textbf{0.810} & \textbf{0.810} & \textbf{0.810}&  \textbf{0.810} & 0.619 \\ \hline
SJETV047 & \textbf{1.0} & \textbf{1.0} & \textbf{1.0} & 0.977 & 0.0 \\ \hline
SJMLL026 & \textbf{1.0} & \textbf{1.0} & 0.930 & 0.892 & 0.0 \\ \hline
SJMLL039 & 0.927 & 0.927 & \textbf{0.960} & 0.907 & 0.014 \\ \hline

\end{tabular}
\caption{Adjusted Rand Index (ARI) for each mutation clustering method on the 14 B-ALL data sets. Bold designates highest (best) ARI. If more than one method achieved the best ARI on a data set, all corresponding methods are bolded. ``Orchard+PAC" and ``Pairtree+PAC" refer to the phylogeny-aware clustering method applied to mutation trees constructed by Orchard and Pairtree, respectively.}
\label{table:ball-clustering-results}
\end{table}

\begin{table}[H]
\centering
\begin{tabular}{|*{2}{c|}}  
\hline
 Data set & Clusters \\ \hline
 SJBALL022609 & 17 \\ \hline
SJBALL022610 & 3 \\ \hline
SJBALL022611 & 8 \\ \hline
SJBALL022612 & 10 \\ \hline
SJBALL022613 & 9 \\ \hline
SJBALL022614 &  7 \\ \hline
SJBALL031 & 5 \\ \hline
SJBALL036 & 11 \\ \hline
SJERG009 & 8 \\ \hline
SJETV010 & 11 \\ \hline
SJETV043 & 8 \\ \hline
SJETV047 & 5 \\ \hline
SJMLL026 &  7 \\ \hline
SJMLL039 &  9  \\ \hline

\end{tabular}
\caption{The number of clusters derived by experts in \cite{dobson_relapse-fated_2020} for the 14 B-ALL data sets. The data can be found here: https://github.com/morrislab/pairtree-experiments}
\label{table:ball-clone-tree-sizes}
\end{table}

\begin{table}[H]
\scriptsize
\begin{tabular}{|*{13}{c|}}  
\hline
\multirow{3}{4em}{} & \multicolumn{12}{|c|}{Simulated data set size (mutations)} \\ \cline{2-13}
 & \multicolumn{2}{|c}{10} & \multicolumn{2}{|c}{30} & \multicolumn{2}{|c}{50} & \multicolumn{2}{|c}{100} & \multicolumn{2}{|c}{150} & \multicolumn{2}{|c|}{200} \\ \cline{2-13}
& \textit{mean} & \textit{std} & \textit{mean} & \textit{std} & \textit{mean} & \textit{std} & \textit{mean} & \textit{std} & \textit{mean} & \textit{std} & \textit{mean} & \textit{std} \\ \hline
CALDER & $<1$ & $<1$ & 99 & 188 & 468 & 201 & NA & NA & NA & NA & NA & NA \\ \hline
Pairtree & 1 & $<1$ & 6 & 1 & 11 & 3 & 46 & 28 & 90 & 28 & 229 & 38 \\ \hline
Orchard (k=1) & $<1$ & $<1$ & $<1$ & 1 & $<1$ & $<1$ & 6 & 6 & 18 & 12 & 34 & 17 \\ \hline
Orchard (k=10) & $<1$ & $<1$ & 1 & 1 & 4 & 2 & 60 & 63 & 198 & 134 & 374 & 182 \\ \hline
\end{tabular}
\caption{Table of wall clock run time means and standard deviations (rounded to the nearest minute) for each method on each problem size for the simulated cancer reconstructions show in Figure \ref{fig:90mutsim_compare}. NA denotes a problem size where a method did not produce a reconstruction. $<1$ represents that the mean or standard deviation of run times for a method on a problem size was less than a minute.}
\label{suptable:simulated-runtimes}
\end{table}

\nolinenumbers
%
%
%

\end{document}